\documentclass[twocolumn]{emulateapj}
\usepackage{ amsmath, color, natbib,graphicx,tablefootnote}
\usepackage{graphics}
\usepackage{subfigure}
\usepackage{hyperref}
\usepackage{graphicx}

\usepackage{amsmath}
\usepackage{natbib}
\usepackage{float}
\newcommand{\CH}[1]{\colhead{#1}}

\newcommand\ii{{\sc ii}}
\newcommand\iii{{\sc iii}}

\begin{document}

\shortauthors{Berg et al.}
\title{CHAOS I. Direct Chemical Abundances for \ion{H}{2} Regions in NGC 628}

\author{Danielle A. Berg\altaffilmark{1,2}, Evan D. Skillman\altaffilmark{1}, Kevin V. Croxall\altaffilmark{3},  
Richard W. Pogge\altaffilmark{3,4}, John Moustakas\altaffilmark{5}, Mara Johnson-Groh\altaffilmark{6}}

\altaffiltext{1}{Center for Gravitation, Cosmology and Astrophysics, Department of Physics, University of Wisconsin Milwaukee,
1900 East Kenwood Boulevard, Milwaukee, WI 53211; bergda@uwm.edu}
\altaffiltext{2}{Minnesota Institute for Astrophysics, University of Minnesota, 116 Church St. SE, Minneapolis, MN 55455;  skillman@astro.umn.edu}
\altaffiltext{3}{Department of Astronomy, The Ohio State University, 140 W. 18th Avenue, Columbus, OH 43202; croxall.5@osu.edu; pogge.1@osu.edu}
\altaffiltext{4}{Center for Cosmology and AstroParticle Physics, The Ohio State University, 191 West Woodruff Avenue, Columbus, OH 43210}
\altaffiltext{5}{Department of Physics and Astronomy, Siena College, 515 Loudon Road, Loudonville, NY 12211; jmoustakas@siena.edu}
\altaffiltext{6}{Department of Physics and Astronomy, University of Victoria, Victoria, British Columbia V8P 1A1, Canada; mara@uvic.ca}


\begin{abstract}

The CHemical Abundances of Spirals (CHAOS) project leverages the combined power of the
Large Binocular Telescope (LBT) with the broad spectral range and sensitivity of the Multi Object Double 
Spectrograph (MODS) to measure ``direct" abundances (based on observations of the temperature-sensitive 
auroral lines) in large samples of \ion{H}{2} regions in spiral galaxies.
We present LBT MODS observations of 62 \ion{H}{2} regions in the nearby spiral galaxy NGC 628,  
with an unprecedentedly large number of auroral lines measurements 
(18 [\ion{O}{3}] $\lambda4363$, 29 [\ion{N}{2}] $\lambda5755$, 40 [\ion{S}{3}]$\lambda6312$, 
and 40 [\ion{O}{2}] $\lambda\lambda7320,7330$ detections) in 45 \ion{H}{2} regions.  
Comparing derived temperatures from multiple auroral line measurements, we find: 
(1) a strong correlation between temperatures based on [\ion{S}{3}] $\lambda6312$ 
and [\ion{N}{2}] $\lambda5755$; and 
(2) large discrepancies for some temperatures based on
[\ion{O}{2}] $\lambda\lambda7320,7330$ and [\ion{O}{3}] $\lambda4363$.
Both of these trends are consistent with other observations in the literature,
yet, given the widespread use and acceptance of [\ion{O}{3}] $\lambda4363$ as a temperature determinant, 
the magnitude of the T[\ion{O}{3}] discrepancies still came as a surprise.
Based on these results, we conduct a uniform abundance analysis prioritizing the temperatures 
derived from [\ion{S}{3}] $\lambda6312$ and [\ion{N}{2}] $\lambda5755$, 
and report the gas-phase abundance gradients for NGC~628. 
Relative abundances of S/O, Ne/O, and Ar/O are constant across the galaxy, 
consistent with no systematic change in the upper IMF over the sampled range in metallicity.
These alpha-element ratios, along with N/O, all show small dispersions  
($\sigma\sim0.1$ dex) over 70\% of the azimuthally averaged radius.  
We interpret these results as an indication that, at a given radius, 
the interstellar medium in NGC~628 is chemically well-mixed.
Unlike the gradients in the nearly temperature-independent relative abundances,
 O/H abundances have a larger intrinsic dispersion of $\sim$ 0.165 dex.  
We posit that this dispersion represents an upper limit to the true
dispersion in O/H at a given radius and that some of that dispersion is
due to systematic uncertainties arising from temperature measurements.

\end{abstract}



\section{INTRODUCTION}\label{sec:intro}

The chemical evolution of galaxies, whereby successive generations of stars enrich 
the interstellar medium (ISM) with products from stellar nucleosynthesis, 
is key to gaining a deeper understanding of galaxy evolution.
\ion{H}{2} regions can be used to study the absolute and relative abundances in the ISM of 
spiral galaxies.
Thus, spiral galaxies in the nearby universe, with low inclinations, 
provide the opportunity for studying and understanding the chemical evolution of galaxies.

Chemical abundances have been widely studied in spiral galaxies using ``strong-line" 
calibrations, where ratios of the strong forbidden lines are used as abundance indicators. 
However, these measurements are only statistical indications of chemical abundance, 
and are limited by the parameter space of the calibration sample
\citep[e.g.,][]{vanzee06a,bresolin07,yin07,bresolin09a,stasinska10,amorin10,berg11}.
Because spiral galaxies exhibit abundance gradients and large ranges in temperature, 
density, and excitation space (among other parameters), ``strong-line" calibrations are 
vulnerable to large systematic uncertainties \citep{kewley08,moustakas10}. 
In contrast, the ``direct" method uses the temperature-sensitive 
ratio of auroral to forbidden emission lines in order to determine the  
electron temperature and subsequent chemical abundances.

Due to the inherent challenges of the direct method, 
detailed direct abundance studies exist for only a handful of spiral galaxies
(e.g., 20 \ion{H}{2} regions with direct abundances in M101, \citet{kennicutt03a};
61 \ion{H}{2} regions with direct abundance in M33, \citet{rosolowsky08}).
This relative dearth of large samples of direct abundances in the ISM of spiral 
galaxies prevents us from understanding key processes in the 
evolution of galaxies and limits our ability to fully understand: 
the chemical abundance gradients in spirals; 
the dispersion in the ISM abundances as a function of radius and environment;
the yields of elements from stellar nucleosynthesis with potential constraints on the stellar initial mass function; 
the chemical evolution of galaxies at high redshift; 
and thus, the chemical enrichment history of the universe.  
In order to place these measurements on the same scale for comparison amongst 
galaxies, reliable and consistent derivations of abundances are needed 
\citep[see, e.g.,][]{moustakas10}.

With recognition that simply measuring a direct abundance is not
the panacea to all \ion{H}{2} region abundance measurement problems,
assembling large, homogeneous datasets of electron temperature measurements
represent our best observational approach to putting all \ion{H}{2} 
region abundance measurements on firmer ground.
Increasing the number of \ion{H}{2} regions observed per galaxy allows for statistical
approaches to questions about
discrepant electron temperature measurements \citep[e.g.,][]{kennicutt03a, binette12}, 
dispersions in abundances at a given radius \citep[e.g.,][]{kennicutt03a, rosolowsky08},
and dispersions as a function of environment \citep[e.g.,][]{kennicutt03a}.

\subsection{The CHAOS Project}
The CHemical Abundances Of Spirals (CHAOS) project\footnotemark[7] was undertaken in order to 
build a large database of high quality \ion{H}{2} region spectra from nearby spiral galaxies using the 
Multi-Object Double Spectrographs \citep[MODS;][]{pogge10} on the Large Binocular Telescope (LBT).
The MODS have been designed to optimally obtain high quality spectra across fields of 
view (6\arcmin$\times$6\arcmin) comparable to the extents of nearby spiral galaxies. 
Furthermore, the LBT and MODS combination provides the balance between sensitivity, 
resolution, and wavelength coverage necessary to measure all emission lines 
relevant to oxygen, nitrogen, and alpha-element abundance determinations and more.
In particular, the optical design of MODS has been optimized for high throughput in the 
red and blue channels to produce high-quality spectrophotometry from the atmospheric 
UV cut-off ($\sim$340 nm) to the silicon detector cutoff at $\sim$1$\mu$m.

The \textit{Spitzer} Infrared Nearby Galaxies Survey \citep[SINGS;][]{kennicutt03b} galaxies are arguably 
the best understood spiral galaxies outside of the Local Group due to having the most complete multi-wavelength observations available.
Due to the \textit{Spitzer} observations and large ancillary data sets, 
these galaxies have resolved 3.6-160 $\mu$m imaging, 5-40 $\mu$m IRS spectroscopy 
of the central regions plus select extra-nuclear \ion{H}{2} regions, and high-quality \ion{H}{1} 
and CO gas maps with detailed 2-D rotation curves. 
Additionally, observations by Herschel under the 
Key Insights on Nearby Galaxies: a Far- Infrared Survey with Herschel project 
\citep[KINGFISH;][]{kennicutt11} provide PACS and SPIRE imaging, 
as well as far-IR spectroscopy of selected \ion{H}{2} regions.
We defined the CHAOS sample by identifying the optimal candidates within the SINGS sample:
we restrict our sample to nearly face-on,  
luminous (M$_{B} < -18$ mag), 
spiral galaxies (Hubble type T $> 0$) accessible from the northern hemisphere 
with the LBT (declination $> -5^{\circ}$), with minimal transit airmasses ($< 1.1$).
For our final target sample, 13 high priority SINGS galaxies were selected.

The collection of a large sample of \ion{H}{2} regions is driven by the need to precisely 
measure the form of the radial gradient, including deviations from simple linear fits, 
and the dispersion in abundance at fixed radius. 
Ideally, roughly 10 \ion{H}{2} regions in approximately 10 independent 
radial bins, or $\sim$ 100 \ion{H}{2} regions per galaxy, would be 
required to achieve the desired precision. 
Given the steep \ion{H}{2} region luminosity function, and the non-random 
distribution of \ion{H}{2} regions in a spiral galaxy, this ideal statistical 
sample is not obtainable for all galaxies, but represents a reasonable goal. 
At present, multiple fields have been observed in 9 of the 13 galaxies,
already more than doubling the number of \ion{H}{2} regions in spiral 
galaxies with sufficient quality spectra to provide accurate measurements 
of the physical conditions (i.e., temperatures, densities, and ionization parameters) 
and absolute and relative chemical abundances. 

\subsection{NGC~628 Properties}
We present LBT observations of \ion{H}{2} regions in the first CHAOS target, NGC~628.
NGC~628 (M~74) is a late-type giant Sc spiral galaxy with a systematic velocity of 656 km s$^{-1}$. 
We adopt a distance of 7.2 Mpc \citep{vandyk06}\footnotemark[8] and an inclination of 
$i\approx5^{\circ}$ \citep{shostak84}, with a resulting scale of 35 pc arcsec$^{-1}$.
The adopted properties for NGC~628 are provided in Table~\ref{tbl1}.
NGC~628 is an excellent target due to its small inclination, 
extended structure, and undisturbed optical profile.
The gas-phase oxygen abundance of NGC~628 has been studied previously  
using long-slit or fiber spectroscopy \citep[][]{talent83,mccall85,zaritsky94,ferguson98,
vanzee98b,bresolin99,castellanos02,moustakas10,gusev12,cedres12,berg13}, 
and integral field spectroscopy \citep[][]{rosales-ortega11,croxall13}.
All together, a combined total of 18 distinct \ion{H}{2} region direct abundance measurements  
exist in the literature for NGC~628 \citep{castellanos02,berg13,croxall13}\footnotemark[9]. 
The CHAOS observations of NGC~628 present a three-fold increase in the total number 
of direct \ion{H}{2} region abundances and allow, for the first time, 
the spatial variations and dispersions in absolute and relative ISM abundances 
to be determined from the direct method.

This paper is organized as follows:
In Section~\ref{data} we describe the LBT observations for NGC~628,
the spectral data processing, the emission line measurements, and
the reddening corrections.
In Section~\ref{sec:phys} we discuss the derivation of the temperatures
and densities of the \ion{H}{2} regions, including the choice of atomic data 
and comparison of the derived temperatures.
Our derived abundances are presented in Section~\ref{sec:abund},
where we introduce our uniform abundance analysis method
and report ionic and absolute abundances for NGC~628.
We further interpret the CHAOS abundance gradients for NGC~628 by considering
electron temperature discrepancies (\S~\ref{sec:discrepant}),
comparing to the abundance gradients in the literature (\S~\ref{sec:comps}),
and assessing the physical basis for dispersions at a given galactocentric radius.
Finally, we summarize our findings in Section~\ref{sec:sum}. 
We also present additional analyses in two Appendices:
The direct oxygen abundance gradient based on [\ion{O}{3}] $\lambda4363$ 
electron temperature measurements is presented in Appendix~\ref{sec:TO3}.
Indicators of temperature discrepancies are presented in Appendix~\ref{sec:indicators}.

\footnotetext[7]{Funding awarded by NSF Grant No. 1109066.}
\footnotetext[8]{\citet{vandyk06} reports a value from supernovae distance determinations.
Current value from the literature for the distance of NGC~628 range from $7.2-9.95$ Mpc.}
\footnotetext[9]{25 observations may be totaled from these three studies, 
however, several originate from the same \ion{H}{2} region complexes. 
Thus, only 18 non-duplicate region observations exist.}


\begin{deluxetable}{lcc}
\tabletypesize{\footnotesize}
\tablewidth{0pt}
\tablecaption{Properties of NGC~628}
\startdata
{Property} 			& {NGC~628}			\\
\hline													
{R.A.}				& {01:36:41.747}		\\
{Dec.}				& {15:47:01.18}			\\
{Type}				& {ScI}				\\
{Adopted D (Mpc)}		& {7.2$\pm$1.0}$^1$ 	\\
{$m_B$ (mag)}			& {9.95}$^2$			\\
{Redshift}				& {0.002192}			\\
{Inclination (degrees)}	& {5}$^3$				\\	
{P.A. (degrees)}			& {12}$^4$			\\
{$R_{25}$ (arcmin)}		& {5.25}$^5$			\\
{$R_{25}$ (kpc)}		& {10.95}				\\
\enddata
\tablecomments{Adopted properties for NGC~628.
Rows 1 and 2 give the RA and Dec of the optical center in units of
hours, minutes, and seconds, and decrees, arcminutes, and arcseconds respectively.
Row 5 lists redshifts taken from the NASA/IPAC Extragalactic Database.
Row 8 gives the optical radius at the $B_{25}$ mag arcsec$^{-2}$ of the system.
Row 9 gives the optical radius of the galaxy given the adopted distance.
References: (1) \citet{vandyk06}; (2) \citet{jlee11}; (3) \citet{shostak84}; (4) \citet{egusa09};
(5) \citet{kendall11} }
\label{tbl1}
\end{deluxetable}



\section{CHAOS SPECTROSCOPIC OBSERVATIONS}\label{data}

\subsection{Observation Plan}
Optical spectra of NGC~628 were acquired with MODS1 on the LBT
as part of the CHAOS project on the UT dates of 2012 October 14 and 16. 
The primary goal was to obtain high signal-to-noise spectra, with detections of  
 intrinsically faint auroral lines (e.g., [\ion{O}{3}] $\lambda$4363, [\ion{N}{2}] $\lambda$5755, 
 [\ion{S}{2}] $\lambda$6312) at a significance of 3$\sigma$ or higher.
 The multi-object mode of MODS, which uses custom-designed, laser-milled slit masks,
 allows spectra of many \ion{H}{2} regions to be obtained simultaneously.
Broad-band and H$\alpha$ continuum-subtracted SINGS images for NGC 628 
\citep{munoz-mateos09} were used to identify target \ion{H}{2} regions, 
as well as alignment stars, and determine accurate astrometry for the masks.
\ion{H}{2} regions were selected to achieve large radial coverage of the optical disk, 
prioritizing knots of highest H$\alpha$ surface brightness.
We observed three multi-slit masks, each containing $\sim$20 slits, 
which spanned the radial and azimuthal extent of the optical disk of NGC~628. 
All \ion{H}{2} region slits are 1.0\arcsec\ wide, but lengths vary between $8-20$\arcsec\
depending on the size of the targeted \ion{H}{2} region and proximity of other slits on the mask.
Blue and red spectra were obtained simultaneously using the G400L (400 lines mm$^{-1}$, R$\sim$1850)
and G670L (250 lines mm$^{-1}$, R$\sim$2300) gratings, respectively.
This combination results in a broad spectral coverage extending from 3200 - 10,000 \AA,
that was linearly re-sampled to 0.5\,\AA\ per pixel and full width half maximum resolution of $\sim$ 2 \AA.

Each mask field was observed for six 1200 second exposures, for a total integration time of two hours.
The masks were designed with slits at a fixed position angle which 
approximated the parallactic angle at the midpoint of the longest possible observation window in a night.
For NGC~628, during the observing run of UTC 2012 October 2$-$17, we selected the parallactic angle of 
PA $= -235.2^{\circ}$ for Fields 1 and 3, and PA $= -55.1^{\circ}$ for Field 2, which is just a rotation of $180^{\circ}$
to allow for the use of different guide stars and thus a shift of the field center.
This, in addition to observing the galaxies at airmasses less than 1.5, 
served to minimize the wavelength-dependent light loss due to 
differential atmospheric refraction \citep{filippenko82}. 
On the nights that NGC~628 was observed, the sky conditions were optimal: clear, low wind, and low humidity. 
Additionally, we aimed for approximately arcsecond seeing. 
Fields 1, 2, and 3 had average seeing values over their 2 hour observation windows of 1.1,
1.25, and 0.85 arcseconds, respectively.  

\begin{figure}
\epsscale{1.2}
\plotone{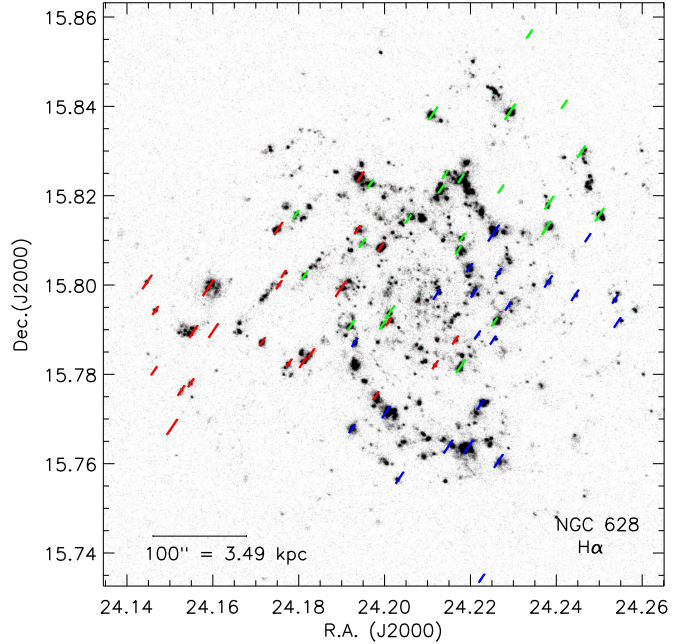}
\caption{Continuum-subtracted H$\alpha$ SINGS image of NGC 628 \citep{munoz-mateos09}.
The footprints of CHAOS slits are overlaid, where red, green, and blue respectively represent the 
Field 1, 2, and 3 slit positions observed at the LBT. 
The slit positions targeted \ion{H}{2} regions, although not always centered in order to maximize
effective usage of mask real estate. 
See Table~\ref{tbl2} for more details.}
\label{fig1}
\end{figure}

Figure~\ref{fig1} shows the H$\alpha$ continuum-subtracted SINGS image for NGC 628 \citep{munoz-mateos09}, 
with slit positions from Table~\ref{tbl2} shown as colored lines for each mask field. 
Throughout this work, all target locations (extraction of \ion{H}{2} region; not slit center) are 
referred to by their offsets in right ascension and declination from the center of NGC~628, 
which is taken to be $01^{h}36^{m}41.7^{s}$ $+15^{\circ}47$\arcmin 01\arcsec .
Slit locations and offsets for all observed \ion{H}{2} regions are listed in Table~\ref{tbl2}.


\begin{deluxetable*}{lcccccc|c|c|c|c|c|}
\tabletypesize{\scriptsize}
\tablewidth{0pt}
\setlength{\tabcolsep}{3pt}
\tablecaption{Logs for NGC~628 LBT Observations\label{tab:slits}}
\tablehead{
\CH{H \ii} 				& \CH{R.A.}	& \CH{Dec.}	& \CH{R$_g$}	  & \CH{R/R$_{25}$} 	& \CH{R$_g$}	& \CH{Offset (R.A., Dec.)}	& \multicolumn{5}{c}{Auroral Line Detections}\\ \cline{8-12} 
\CH{Region}			& \CH{(2000)} 	& \CH{(2000)} 	& \CH{(arcsec)}	  & \CH{}			& \CH{(kpc)} 	& \CH{(arcsec)}			& \CH{[O \iii]}	& \CH{[N \ii] }	& \CH{[S \iii]} & \CH{[O \ii]} & \CH{[S \ii]}}
\startdata
Total Detections:		& {}			& {}			& {}				& {}		&& {}		 		& 18			& 29			& 40			& 40			& 0			\\				
\hline
NGC628-12.7+2.1 		& 1:36:40.92	& 15:47:02.06	& 12.037	& 0.038	& 0.42	& -12.7,+2.1		& {}			& {}			& {}			& {}			& {}			\\	
NGC628+38.6-17.8 		& 1:36:44.47	& 15:46:42.19	& 43.863	& 0.139	& 1.53	& +38.6,-17.8		& {}			& {}			& {}			& {}			& {}			\\	
NGC628-48.8+0.7 		& 1:36:38.42	& 15:47:00.72	& 48.194	& 0.153	& 1.68	& -48.8,+0.7		& {}			& {}			& {}			& {}			& {}			\\	
NGC628+38.9-26.7 		& 1:36:44.49	& 15:46:33.32	& 48.628	& 0.154	& 1.69	& +38.9,-26.7		& {}			& {}			& {}			& {}			& {}			\\ 	
NGC628-44.7+27.3 		& 1:36:38.70	& 15:47:27.34	& 51.311	& 0.163	& 1.79	& -44.7,+27.3		& {}			& {}			& {}			& {}			& {}			\\	
NGC628-33.2+45.8 		& 1:36:39.50	& 15:47:45.80	& 55.297	& 0.176	& 1.93	& -33.2,+45.8		& {}			& {}			& {}			& {}			& {}			\\	
NGC628-30.7-46.5 		&  1:36:39.67	& 15:46:13.48	& 56.337	& 0.179	& 1.97	& -30.7,-46.5		& {}			& {}			& {}			& {}			& {}			\\	
NGC628+48.0-32.0 		& 1:36:45.13	& 15:46:27.98	& 59.216	& 0.188	& 2.07	& +48.0,-32.0		& {}			& {}			& {}			& {}			& {}			\\	
NGC628-35.9+57.7 		& 1:36:39.31	& 15:47:57.72	& 66.695	& 0.212	& 2.33	& -35.9,+57.7		& {}			& \checkmark	& \checkmark	& {}			& {}			\\	
NGC628-55.5-40.4 		& 1:36:37.96	& 15:46:19.64	& 68.798	& 0.218	& 2.40	& -55.5,-40.4		& {}			& {}			& {}			& {}			& {}			\\	
NGC628+49.8+48.7 		& 1:36:45.25	& 15:47:48.72	& 69.474	& 0.221	& 2.43	& +49.8,+48.7		& {}			& \checkmark	& \checkmark	& \checkmark	& {}			\\	
NGC628-8.3-72.8 		& 1:36:41.23	& 15:45:47.18	& 74.386	& 0.236	& 2.59	& -8.3,-72.8		& {}			& {}			& {}			& {}			& {}			\\	
NGC628-73.1-27.3 		& 1:36:36.73	& 15:46:32.68	& 77.973	& 0.248	& 2.72	& -73.1,-27.3		& {}			& {}			& \checkmark	& \checkmark	& {}			\\	
NGC628-76.2+22.9 		& 1:36:36.52	& 15:47:22.91	& 78.816	& 0.250	& 2.75	& -76.2,+22.9		& {}			& \checkmark	& \checkmark	& \checkmark	& {}			\\	
NGC628+18.5+79.0 		& 1:36:43.08	& 15:48:19.02	& 80.178	& 0.255	& 2.80	& +18.5,+79.0		& {}			& {}			& {}			& {}			& {}			\\	
NGC628-36.8-73.4 		& 1:36:39.25	& 15:45:46.65	& 82.827	& 0.263	& 2.89	& -36.8,-73.4		& {}			& \checkmark	& {}			& \checkmark	& {}			\\	
NGC628-73.1-44.8 		& 1:36:36.74	& 15:46:15.18	& 85.853	& 0.273	& 3.00	& -73.1,-44.8		& {}			& {}			& {}			& {}			& {}			\\	
NGC628+68.5+53.4 		& 1:36:46.54	& 15:47:53.35	& 86.834	& 0.276	& 3.04	& +68.5,+53.4		& {}			& \checkmark	& \checkmark	& \checkmark	& {}			\\	
NGC628-87.5-10.8 		& 1:36:35.74	& 15:46:49.16	& 87.870	& 0.279	& 3.07	& -87.5,-10.8		& {}			& {}			& {}			& {}			& {}			\\	
NGC628+81.6-32.3 		& 1:36:47.45	& 15:46:27.69	& 89.257	& 0.283	& 3.11	& +81.6,-32.3		& {}			& {}			& \checkmark	& {}			& {}			\\	
NGC628-68.5+61.7 		& 1:36:37.05	& 15:48:01.74	& 91.116	& 0.289	& 3.18	& -68.5,+61.7		& {}			& \checkmark	& \checkmark	& \checkmark	& {}			\\	
NGC628+76.9-49.6 		& 1:36:47.13	& 15:46:10.42	& 93.074	& 0.295	& 3.25	& +76.9,-49.6		& {}			& \checkmark	& \checkmark	& \checkmark	& {}			\\	
NGC628+94.2+4.0 		& 1:36:48.33	& 15:47:03.99	& 95.361	& 0.303	& 3.33	& +94.2,+4.0		& {}			& {}			& {}			& {}			& {}			\\	
NGC628+75.4+66.0 		& 1:36:47.02	& 15:48:06.01	& 100.177	& 0.318	& 3.50	& +75.4,+66.0		& {}			& {}			& {}			& {}			& {}			\\	
NGC628-13.1+107.5 	& 1:36:40.89	& 15:48:47.50	& 107.076	& 0.340	& 3.73	& -13.1,+107.5		& {}			& \checkmark	& \checkmark	& \checkmark	& {}			\\	
NGC628+53.5-104.0 	& 1:36:45.51	& 15:45:16.04	& 118.496	& 0.376	& 4.13	& +53.5,-104.0		& {}			& \checkmark	& {}			& \checkmark	& {}			\\	
NGC628-35.7+119.6 	& 1:36:39.33	& 15:48:59.59	& 123.566	& 0.392	& 4.31	& -35.7,+119.6		& {}			& \checkmark	& \checkmark	& \checkmark	& {}			\\	
NGC628-20.3+124.6 	& 1:36:40.40	& 15:49:04.62	& 125.033	& 0.397	& 4.36	& -20.3,+124.6		& {}			& \checkmark	& \checkmark	& \checkmark	& {}			\\	
NGC628-59.6-111.6 		& 1:36:37.67	& 15:45:08.43	& 127.196	& 0.404	& 4.44	& -59.6,-111.6		& {}			& \checkmark	& \checkmark	& \checkmark	& {}			\\	
NGC628+61.2+113.5 	& 1:36:46.04	& 15:48:53.45	& 128.290	& 0.407	& 4.48	& +61.2,+113.5		& {}			& \checkmark	& \checkmark	& \checkmark	& {}			\\	
NGC628-128.4+13.1 	& 1:36:32.90	& 15:47:13.09	& 128.700	& 0.409	& 4.50	& -128.4,+13.1		& {}			& {}			& {}			& {}			& {}			\\	
NGC628+42.6-120.7 	& 1:36:44.75	& 15:44:59.30	& 129.490	& 0.411	& 4.52	& +42.6,-120.7		& {}			& \checkmark	& \checkmark	& \checkmark	& {}			\\	
NGC628+131.9+18.5 	& 1:36:50.94	& 15:47:18.53	& 134.275	& 0.426	& 4.69	& +131.9,+18.5		& \checkmark	& \checkmark	& \checkmark	& \checkmark	& {}			\\	
NGC628+71.2+121.3 	& 1:36:46.74	& 15:49:01.32	& 140.117	& 0.445	& 4.90	& +71.2,+121.3		& {}			& {}			& {}			& {}			& {}			\\	
NGC628+125.4-62.4 	& 1:36:50.49	& 15:45:57.61	& 141.767	& 0.450	& 4.95	& +125.4,-62.4		& {}			& {}			& \checkmark	& \checkmark	& {}			\\	
NGC628-130.9+71.8 	& 1:36:32.73	& 15:48:11.80	& 148.566	& 0.472	& 5.18	& -130.9,+71.8		& {}			& \checkmark	& \checkmark	& \checkmark	& {}			\\	
NGC628+131.7-70.2 	& 1:36:50.93	& 15:45:49.80	& 151.024	& 0.479	& 5.27	& +131.7,-70.2		& {}			& {}			& \checkmark	& \checkmark	& {}			\\	
NGC628+151.0+22.3 	& 1:36:52.26	& 15:47:22.33	& 153.790	& 0.488	& 5.37	& +151.0,+22.3		& \checkmark	& \checkmark	& \checkmark	& \checkmark	& {}			\\	
NGC628-157.9-0.3 		& 1:36:30.86	& 15:46:59.71	& 157.687	& 0.501	& 5.51	& -157.9,-0.3		& {}			& \checkmark	& \checkmark	& \checkmark	& {}			\\	
NGC628-24.5-155.6 		& 1:36:40.10	& 15:44:24.39	& 158.586	& 0.503	& 5.54	& -24.5,-155.6		& {}			& \checkmark	& \checkmark	& \checkmark	& {}			\\	
NGC628-129.8+94.7 	& 1:36:32.81	& 15:48:34.73	& 159.867	& 0.508	& 5.58	& -129.8,+94.7		& {}			& \checkmark	& \checkmark	& \checkmark	& {}			\\	
NGC628+158.5+9.2 		& 1:36:52.78	& 15:47:09.20	& 160.025	& 0.508	& 5.58	& +158.5,+9.2		& {}			& {}			& {}			& \checkmark	& {}			\\	
NGC628+78.6-137.6 	& 1:36:47.24	& 15:44:42.37	& 160.143	& 0.508	& 5.59	& +78.6,-137.6		& {}			& {}			& {}			& \checkmark	& {}			\\	
NGC628+140.3+82.0 	& 1:36:51.52	& 15:48:21.99	& 162.919	& 0.517	& 5.68	& +140.3,+82.0		& {}			& {}			& \checkmark	& {}			& {}			\\	
NGC628-42.8-158.2 		& 1:36:38.84	& 15:44:21.79	& 164.834	& 0.523	& 5.75	& -42.8,-158.2		& \checkmark	& \checkmark	& \checkmark	& \checkmark	& {}			\\	
NGC628+147.9-71.8 	& 1:36:52.05	& 15:45:48.18	& 166.229	& 0.528	& 5.80	& +147.9,-71.8		& {}			& \checkmark	& {}			& \checkmark	& {}			\\	
NGC628+163.5+64.4 	& 1:36:53.13	& 15:48:04.34	& 176.489	& 0.560	& 6.16	& +163.5,+64.4		& {}			& \checkmark	& \checkmark	& \checkmark	& {}			\\	
NGC628-4.5+185.6 		& 1:36:41.49	& 15:50:05.63	& 184.519	& 0.586	& 6.44	& -4.5,+185.6		& {}			& \checkmark	& \checkmark	& \checkmark	& {}			\\	
NGC628+176.7-50.0 	& 1:36:54.04	& 15:46:09.96	& 185.425	& 0.589	& 6.47	& +176.7,-50.0		& {}			& {}			& \checkmark	& \checkmark	& {}			\\	
NGC628-76.2-171.8 		& 1:36:36.52	& 15:44:08.22	& 188.715	& 0.599	& 6.59	& -76.2,-171.8		& \checkmark	& {}			& \checkmark	& \checkmark	& {}			\\	
NGC628+31.6-191.1 	& 1:36:43.99	& 15:43:48.85	& 195.131	& 0.619	& 6.81	& +31.6,-191.1		& \checkmark	& \checkmark	& \checkmark	& \checkmark	& {}			\\	
NGC628-200.6-4.2 		& 1:36:27.90	& 15:46:55.81	& 200.634	& 0.637	& 7.00	& -200.6,-4.2		& \checkmark	& \checkmark	& \checkmark	& \checkmark	& {}			\\	
NGC628-184.7+83.4 	& 1:36:29.00	& 15:48:23.36	& 202.182	& 0.642	& 7.06	& -184.7,+83.4		& \checkmark	& \checkmark	& \checkmark	& \checkmark	& {}			\\	
NGC628-206.5-25.7 		& 1:36:27.49	& 15:46:34.31	& 208.222	& 0.661	& 7.27	& -206.5,-25.7		& \checkmark	& {}			& \checkmark	& \checkmark	& {}			\\	
NGC628-90.1+190.2 	& 1:36:35.56	& 15:50:10.17	& 209.327	& 0.665	& 7.31	& -90.1,+190.2		& \checkmark	& \checkmark	& \checkmark	& \checkmark	& {}			\\	
NGC628-168.2+150.8 	& 1:36:30.14	& 15:49:30.77	& 225.204	& 0.715	& 7.86	& -168.2,+150.8 	& \checkmark	& {}			& \checkmark	& \checkmark	& {}			\\	
NGC628+232.7+6.6 		& 1:36:57.92	& 15:47:06.58	& 234.411	& 0.744	& 8.19	& +232.7,+6.6		& \checkmark	& {}			& \checkmark	& \checkmark	& {}			\\	
NGC628+237.6+3.0 		& 1:36:58.26	& 15:47:02.93	& 239.203	& 0.759	& 8.35	& +237.6,+3.0		& \checkmark	& {}			& \checkmark	& \checkmark	& {}			\\	
NGC628+254.3-42.8 	& 1:36:59.42	& 15:46:17.11	& 259.853	& 0.825	& 9.07	& +254.3,-42.8		& \checkmark	& {}			& \checkmark	& {}			& {}			\\	
NGC628+252.1-92.1 	& 1:36:59.26	& 15:45:27.85	& 270.527	& 0.859	& 9.45	& +252.1,-92.1		& \checkmark	& \checkmark	& \checkmark	& {}			& {}			\\	
NGC628+261.9-99.7 	& 1:36:59.94	& 15:45:20.24	& 282.451	& 0.897	& 9.89	& +261.9,-99.7		& \checkmark	& {}			& {}			& {}			& {}			\\	
NGC628+265.2-102.2 	& 1:37:00.17	& 15:45:17.76	& 286.416	& 0.909	& 10.00	& +265.2,-102.2 	& \checkmark	& {}			& {}			& {}			& {}			\\	
NGC628+289.9-17.4 	& 1:37:01.89	& 15:46:42.55	& 292.382	& 0.928	& 10.20	& +289.9,-17.4		& \checkmark	& {}			& \checkmark	& \checkmark	& {}			\\	
NGC628+298.4+12.3 	& 1:37:02.47	& 15:47:12.26	& 300.442	& 0.954	& 10.49	& +298.4,+12.3		& \checkmark	& {}			& \checkmark	& \checkmark	& {}			\\	
\enddata
\tablecomments{Observing logs for \ion{H}{2} regions observed in NGC~628  
using MODS of the LBT on the UT dates of 14 and 16 October 2012.
Each field was observed over an integrated exposure time of 1200 s on clear nights,
with on average $\sim1\arcsec$ seeing, and airmasses less than 1.5. 
\ion{H}{2} region IDs are listed in Column 1, labeled according to their offsets, 
and ordered by increasing galactocentric radius.
The right ascension and declination of the individual \ion{H}{2} regions are given in units of
hours, minutes, and seconds, and decrees, arcminutes, and arcseconds respectively.
The \ion{H}{2} region distances from the center of the galaxy in arcseconds, fraction of $R_{25}$,
and in kpc are listed in the Columns 4-6.
Additionally, the offsets from galaxy center R.A. and Dec.
are given in arcseconds in Column 7. Finally, Columns 8-12 highlight which regions
have [\ion{O}{3}] $\lambda4363$, [\ion{N}{2}] $\lambda5755$, [\ion{S}{3}] $\lambda6312$,
\ion{O}{2}] $\lambda\lambda7320,7330$, and \ion{S}{2}] $\lambda\lambda4068,4076$
auroral lines detections at the 4$\sigma$ significance level. }
\label{tbl2}
\end{deluxetable*}


\subsection{Spectral Reductions}\label{sec:reduct}

Spectra are reduced and analyzed using the beta-release version of the MODS reduction pipeline\footnotemark[10].
Bias subtraction and flat-fielding were performed using python scripts 
designed to accommodate the interlaced data read-out. 
The pixel flat is assembled from slitless lamp flats of the incandescent continuum lamp. 
Slit definitions, wavelength calibration, and creation of variance images are performed using 
new tools written to work within the XIDL\footnotemark[11] reduction package. 
Slit locations on two-dimensional images are defined by mapping
laser-milled mask positions and wavelength to pixel coordinates on the CCD.
Wavelength calibrations are very stable, modulo a small flexure shift of order a few pixels. 
Spectra of Hg + Ar, Ne, and Xe + Kr lamps are taken through the slit masks during a calibration 
sequence performed at the beginning of the observing run. 
Because the MODS spectrographs are remarkably stable, the optical path is mapped
by using the images of lamps through specially designed pinhole masks.
With this information, we generate a transformation from an X/Y location on the mask to a pixel on the detector as a function of wavelength,
and this geometric transformation allows the construction of an initial trace.
The initial spectral traces are corrected for atmospheric refraction using the hour angle and 
airmass corresponding to the mid-point of the observation.
This extraction technique was designed for and is effective for use with very faint spectra where continua are faint or not visible.
No slit-losses are observed above 3500 \AA\, and so no correction is made to this end. 

\footnotetext[10]{The MODS reduction pipeline was developed by Kevin Croxall with funding from NSF Grant AST-1108693. 
Details can be found at -- http://www.astronomy.ohio-state.edu/MODS/Software/modsIDL/.}
\footnotetext[11]{http://www.ucolick.org/\~xavier/IDL/.}

Due to the large angular size of NGC 628, the diffuse ionized medium fills the majority of each field of view. 
In order to facilitate accurate sky subtraction, additional ``sky'' slits are placed well away 
from both \ion{H}{2} regions and the main body of the galaxy. 
Due to the off-axis nature of the sky-slits, they are more affected by aberration than slits targeting \ion{H}{2} regions. 
A sky frame was created by fitting a two-dimensional B-spline to the background 
sky in the two-dimensional spectra, i.e., each slit is fit independently with a local sky. 
The continuum flux from the sky-slit is scaled to the background continuum measured locally in each slit. 
Subsequently, the background under the strong nebular lines is interpolated from the scaled 
sky-slit to create the final sky-model that is subtracted.

\begin{figure*}
\epsscale{1.2}
\centering
{\textbf{NGC~628-42.8-158.2}\par\medskip}
\vspace*{-0.5cm}
\plotone{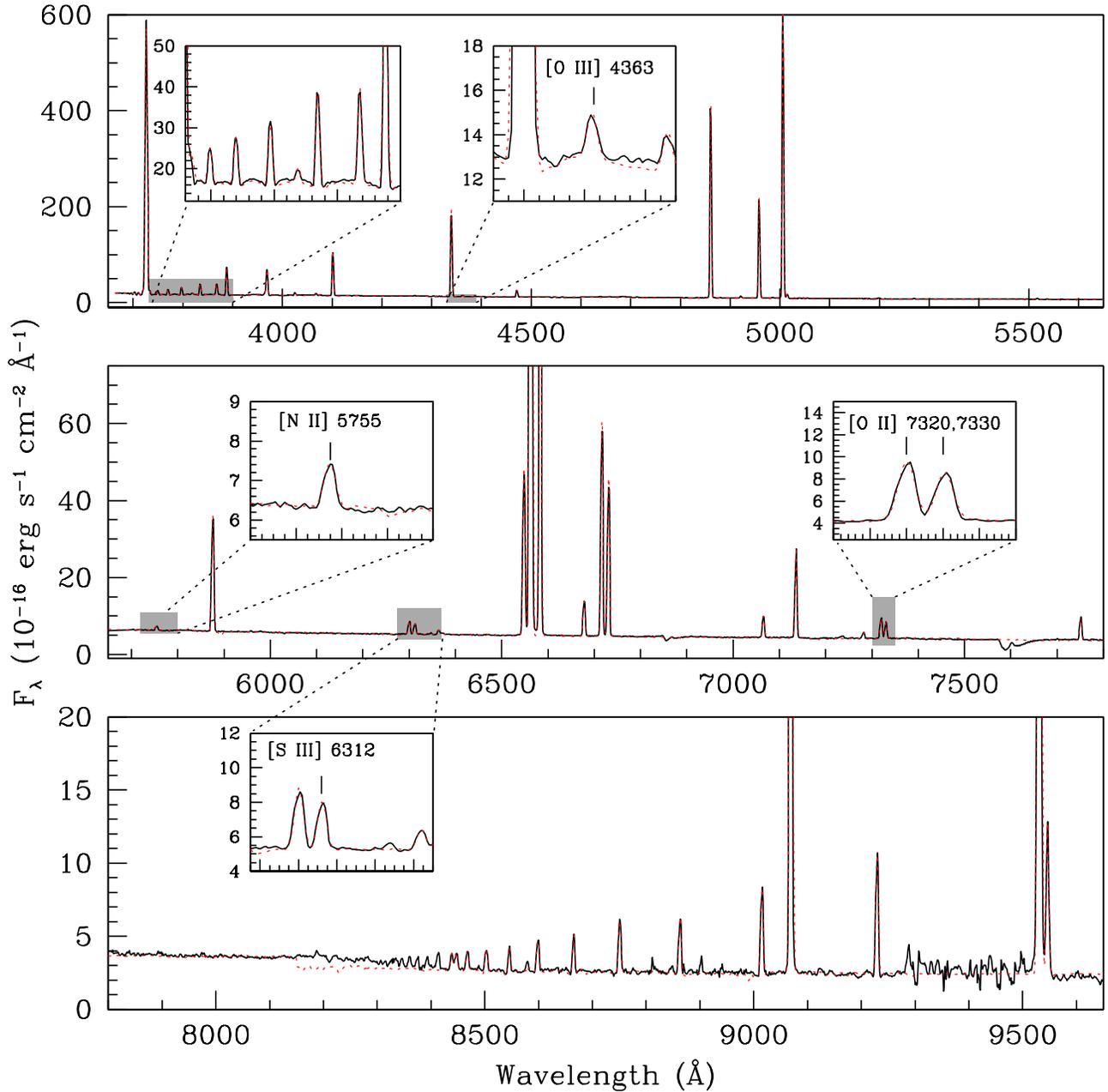}
\caption{Sample LBT spectrum of an \ion{H}{2} region in 
NGC 628 with auroral line detections at a strength of 3$\sigma$ or greater.
The spectrum is broken up into three sections:
The top panel is the blue region, where the high quality of the spectrum is evident by 
the high signal to noise Balmer series (highlighted in the expanded box).
The middle panel corresponds to the middle portion of the spectrum and the bottom
panel to the red region of the spectrum.
The strength of the auroral lines used in this paper [\ion{O}{3}] $\lambda4636$, [\ion{N}{2}] $\lambda5755$, 
[\ion{S}{3}] $\lambda6312$, and [\ion{O}{2}] $\lambda\lambda7320,7330$) are highlighted in the corresponding expanded boxes.
}
\label{fig2}
\end{figure*}

One-dimensional spectra are corrected for atmospheric extinction and flux calibrated based on 
observations of CALSPEC flux standard stars \citep{bohlin10}. 
The flux standards are observed on each night science data are obtained. 
We confirm the high level of fidelity across the spectra by comparing the continuum levels of the red and blue spectra
and find that corrections for matching the two sides are on the order of $\sim1\%$.
An example flux-calibrated spectrum is shown in Figure~\ref{fig2}.


\subsection{Emission Line Measurements}\label{sec:iraf}
Emission line strengths are measured using various existing tools described in \citet{croxall13}.
The underlying continuum of our MODS1 spectra are modeled using the 
STARLIGHT\footnotemark[12] spectral synthesis code \citep{fernandes05}. 
We fit \citet{bruzual13} models to the blue continuum red-ward of the [\ion{O}{2}] $\lambda$3727 line, 
after masking nebular emission lines and Wolf-Rayet features, using a \citet{calzetti00} reddening law. 
Subsequently, emission lines are fit simultaneously with stellar and nebular continuum components.
When automatically fitting emission lines, we initially assume these lines are well described by a Gaussian profile.  
Once continuum levels and line locations are measured, 
the final emission line values are computed by integrating the flux over each profile width.
This step is important with respect to the high resolution of MODS data, because nebular structure can be seen 
in the emission line profiles such that they are not well-characterized by a Gaussian function.
To ensure that noise spikes and multiple features within the same line are not mis-identified and fit, 
emission lines are required to have the same bulk kinematics and FWHM. 
We validate the automated line fits by measuring integrated fluxes by hand using the 
{\tt SPLOT} routine within IRAF\footnotemark[13] for all 62 extracted one-dimensional spectra.  
This exercise confirms the reliability of the automated method.   

\footnotetext[12]{http://astro.ufsc.br/starlight/.} 
\footnotetext[13]{IRAF is distributed by the National Optical Astronomical Observatories.}

In the past, standard errors associated with a given line strength often
followed a relatively simple, generalized prescription \citep[e.g.,][]{skillman94}.
Here, the error associated with each line detection is a combination of the measured variance 
extracted from the two-dimensional variance image, the error associated with the sensitivity function, 
the Poisson noise in the continuum, the read noise of the detector, sky noise, 
a flat fielding calibration error, the error in the continuum placement, 
and uncertainty associated with reddening.
For weak lines, the uncertainty is dominated by error from the continuum subtraction,
whereas the lines with flux measurements much stronger than the rms noise of the continuum
have errors dominated by error in flux calibration and reddening.
To account for the inherent uncertainty in the flux calibration based on standard star observations \citep{oke90}, 
a minimum uncertainty of 2\% is added in quadrature to the emission line flux uncertainties.
Note that this is a conservative estimate as the use of \textit{HST} has allowed improved
calibrations of flux standards in CALSPEC \citep{bohlin10}.
The minimum 2\% emission line uncertainty forces a floor of $\sim$ 50 K uncertainty 
for the electron temperature calculations.

In this study, we focus on the most reliable spectra wherein at least one temperature-sensitive 
auroral line or doublet (i.e., [\ion{O}{3}] $\lambda$4363, [\ion{N}{2}] $\lambda$5755, [\ion{S}{3}] 
$\lambda$6312, or [\ion{O}{2}] $\lambda\lambda$7320,7330) is detected at a line strength $>3\sigma$. 
Note that this criterion is enforced to ensure that, despite best spectral reduction practices,
contamination from other lines, continuum noise, remnant dichroic effects, etc. do not 
lead to false auroral line detections. 
In addition to this precaution, each auroral line is examined by eye to ensure no valid 
detections are excluded and no suspicious detections pass the cut.
By this criterion, out of 62 observed \ion{H}{2} regions, we detect at least one auroral line in 45 spectra.

\subsection{Reddening Corrections}\label{sec:redcor}
The relative intensities of the Balmer lines are nearly independent of both 
density and temperature, so they can be used to solve for the reddening. 
The spectra are de-reddened using an iterative application of the reddening 
law of \citet{odonnell94}, parametrized by $A_{V}=3.1\ E(B-V)$.
An initial reddening estimate is determined from the Balmer line ratios assuming 
a temperature of $T_e = 10^4$ K and a density of  $n_e = 10^2$ cm$^{-3}$.
Using this reddening, we determine an initial estimate of the electron temperature,
which is an error weighted average of the electron temperatures calculated from the auroral lines.
This new electron temperature is then used iteratively to modify the input assumptions for the reddening code
until the average electron temperature changes by less than 20 K. 
The final reddening estimate is an error weighted average of the individual reddening values determined 
from the H$\delta$/H$\beta$, H$\gamma$/H$\beta$, and H$\alpha$/H$\beta$ ratios.
As the lines are fit simultaneously with stellar population models, further corrections
for underlying Balmer absorption are unnecessary. 

Following \citet{lee04}, the reddening value can be converted to the 
logarithmic extinction at $H\beta$ as C(H$\beta) = 1.43 E(B-V)$.
Because of its high Galactic latitude, the foreground extinction for NGC~628 is expected to be low,
with an approximate value of C(H$\beta$) = 0.10 \citep{schlafly11}.
For NGC~628 we find a range in C(H$\beta$) of $\sim0.0$ to 0.6.
Because the MODS spectra extend out to the 1 \micron\ cutoff, we observe the Paschen series and so 
perform an independent calculation of the reddening parameter which confirms the Balmer determinations.   

Reddening corrected line intensities measured from the observed 
\ion{H}{2} regions in the target fields are reported in Table~\ref{tbl3} (the full table is available online).
Various optical emission line ratios can be used to distinguish 
emission regions which are excited by different mechanisms
\citep[e.g.,][]{baldwin81}. 
All 45 regions in the CHAOS sample clearly lie within the typical ranges observed for \ion{H}{2} regions,
ruling out contributions from active galactic nuclei and supernovae remnants, 
and allowing the standard direct analysis of physical conditions in \ion{H}{2} regions. 


\begin{deluxetable*}{lccccccc}
\tabletypesize{\scriptsize}
\setlength{\tabcolsep}{0.04in} 
\tablecaption{ Emission-Line Intensities and Equivalent Widths for LBT Observations of \ion{H}{2} regions in NGC~628}
\tablewidth{0pt}
\tablehead{
\CH{} & \multicolumn{7}{c}{$I(\lambda)/I(\mbox{H}\beta)$} }
\startdata
{Ion}                       			& {-35.9+57.7}      	& {+49.8+48.7}      	& {-73.1-27.3}      	& {-76.2+22.9}      	& {-36.8-73.4}      	& {+68.5+53.4}      	& {+81.6-32.3}       	  \\
\hline
{H14 $\lambda$3721}         	& 0.020$\pm$0.001  	& 0.022$\pm$0.001  	& 0.012$\pm$0.001  	& 0.018$\pm$0.003  	& 0.015$\pm$0.001  	& 0.012$\pm$0.002  	& 0.015$\pm$0.001    \\
{[O~\ii]~$\lambda$3726}     	& 0.475$\pm$0.009  	& 0.694$\pm$0.018  	& 0.492$\pm$0.010  	& 0.507$\pm$0.022  	& 0.472$\pm$0.009  	& 0.626$\pm$0.013  	& 0.352$\pm$0.007    \\
{[O~\ii]~$\lambda$3728}     	& 0.565$\pm$0.011  	& 0.959$\pm$0.020  	& 0.642$\pm$0.013  	& 0.658$\pm$0.022  	& 0.778$\pm$0.016  	& 0.784$\pm$0.016  	& 0.580$\pm$0.012    \\
{H13 $\lambda$3734}         	& 0.025$\pm$0.001  	& 0.027$\pm$0.001  	& 0.015$\pm$0.001  	& 0.022$\pm$0.003  	& 0.019$\pm$0.001  	& 0.015$\pm$0.003  	& 0.019$\pm$0.001    \\
{H12 $\lambda$3750}         	& 0.034$\pm$0.002  	& 0.037$\pm$0.003  	& 0.015$\pm$0.001  	& 0.027$\pm$0.008  	& 0.015$\pm$0.003  	& 0.019$\pm$0.006  	& 0.017$\pm$0.001    \\
{H11 $\lambda$3770}         	& 0.045$\pm$0.001  	& 0.049$\pm$0.003  	& 0.035$\pm$0.001  	& 0.036$\pm$0.008  	& 0.027$\pm$0.003  	& 0.023$\pm$0.006  	& 0.036$\pm$0.001    \\
{H10 $\lambda$3797}         	& 0.054$\pm$0.001  	& 0.058$\pm$0.003  	& 0.034$\pm$0.001  	& 0.048$\pm$0.008  	& 0.041$\pm$0.003  	& 0.031$\pm$0.005  	& 0.041$\pm$0.001    \\
{He~I~$\lambda$3819}       	& 0.007$\pm$0.001  	& 0.016$\pm$0.003  	& 0.003$\pm$0.001  	& $<\sim$0.022        	& $<\sim$~0.007      	& 0.024$\pm$0.005  	& 0.010$\pm$0.001    \\
{H9 $\lambda$3835}          	& 0.070$\pm$0.001  	& 0.082$\pm$0.003  	& 0.064$\pm$0.001  	& 0.068$\pm$0.007  	& 0.079$\pm$0.002  	& 0.075$\pm$0.005  	& 0.067$\pm$0.001    \\
{[Ne~\iii]~$\lambda$3868}   	& $<\sim$~0.004      	& 0.012$\pm$0.002  	& 0.006$\pm$0.002  	& $<\sim$~0.023      	& $<\sim$~0.007      	& 0.017$\pm$0.003  	& $<\sim$~0.003        \\
{He~I~$\lambda$3888}        	& 0.050$\pm$0.001  	& 0.052$\pm$0.003  	& 0.082$\pm$0.002  	& 0.046$\pm$0.007  	& 0.094$\pm$0.002  	& 0.080$\pm$0.005  	& 0.085$\pm$0.002    \\
{H8 $\lambda$3889}          	& 0.105$\pm$0.003  	& 0.111$\pm$0.005  	& 0.065$\pm$0.002  	& 0.093$\pm$0.015  	& 0.079$\pm$0.005  	& 0.060$\pm$0.010  	& 0.080$\pm$0.002    \\
{He~I~$\lambda$3964}        	& 0.008$\pm$0.001  	& 0.013$\pm$0.002  	& 0.008$\pm$0.001  	& $<\sim$~0.021      	& $<\sim$~0.006      	& 0.013$\pm$0.004  	& $<\sim$~0.003        \\
{[Ne~\iii]~$\lambda$3967}   	& \ldots            		& $<\sim$~0.009      	& 0.069$\pm$0.008  	& $<\sim$~0.019      	& $<\sim$~0.056      	& $<\sim$~0.076      	& $<\sim$~0.020        \\
{H7 $\lambda$3970}          	& 0.156$\pm$0.004  	& 0.163$\pm$0.007  	& 0.098$\pm$0.003  	& 0.138$\pm$0.022  	& 0.116$\pm$0.007  	& 0.087$\pm$0.015  	& 0.120$\pm$0.003    \\
{[Ne~\iii]~$\lambda$4011}   	& \ldots            		& $<\sim$~0.004      	& \ldots            		& \ldots            		& $<\sim$~0.005      	& \ldots            		& 0.008$\pm$0.001    \\
{He~I~$\lambda$4026}        	& 0.011$\pm$0.001  	& 0.019$\pm$0.002  	& 0.010$\pm$0.001  	& $<\sim$~0.014      	& 0.011$\pm$0.001  	& $<\sim$~0.008      	& 0.017$\pm$0.001    \\
{[S~\ii]~$\lambda$4068}     	& 0.007$\pm$0.001  	& 0.011$\pm$0.001  	& 0.011$\pm$0.001  	& $<\sim$~0.012      	& $<\sim$~0.004      	& 0.014$\pm$0.003  	& $<\sim$~0.002        \\
{[S~\ii]~$\lambda$4076}     	& 0.003$\pm$0.001  	& 0.008$\pm$0.001  	& 0.006$\pm$0.001  	& $<\sim$~0.012      	& \ldots            		& $<\sim$~0.008      	& \ldots             		  \\
{H$\delta$ $\lambda$4101}   	& 0.247$\pm$0.005  	& 0.271$\pm$0.005  	& 0.275$\pm$0.006  	& 0.253$\pm$0.005  	& 0.276$\pm$0.006  	& 0.288$\pm$0.006  	& 0.250$\pm$0.005    \\
{He~I~$\lambda$4120}        	& 0.003$\pm$0.001  	& 0.012$\pm$0.001  	& 0.004$\pm$0.001  	& $<\sim$~0.012      	& 0.004$\pm$0.001  	& \ldots            		& 0.010$\pm$0.001    \\
{He~I~$\lambda$4143}        	& 0.004$\pm$0.001  	& $<\sim$~0.003      	& \ldots            		& $<\sim$~0.011      	& $<\sim$~0.003      	& $<\sim$~0.006      	& 0.002$\pm$0.001    \\
{H$\gamma$ $\lambda$4340} 	& 0.466$\pm$0.009  	& 0.464$\pm$0.009  	& 0.481$\pm$0.010  	& 0.465$\pm$0.009  	& 0.488$\pm$0.010  	& 0.478$\pm$0.010  	& 0.474$\pm$0.009    \\
{[O~\iii]~$\lambda$4363}    	& $<\sim$~0.004      	& $<\sim$~0.008      	& $<\sim$~0.004      	& $<\sim$~0.011      	& $<\sim$~0.004      	& $<\sim$~0.006      	& $<\sim$~0.003        \\
{He~I~$\lambda$4387}        	& 0.006$\pm$0.001  	& 0.008$\pm$0.001  	& 0.007$\pm$0.001  	& $<\sim$~0.010      	& 0.011$\pm$0.001  	& 0.014$\pm$0.002  	& 0.010$\pm$0.001    \\
{He~I~$\lambda$4471}        	& 0.024$\pm$0.001  	& 0.038$\pm$0.001  	& 0.021$\pm$0.001  	& 0.019$\pm$0.003  	& 0.018$\pm$0.001  	& 0.022$\pm$0.002  	& 0.028$\pm$0.001    \\
{[Fe~\iii]~$\lambda$4658}   	& 0.002$\pm$0.001  	& 0.005$\pm$0.001  	& 0.006$\pm$0.001  	& $<\sim$~0.009      	& \ldots            		& \ldots            		& \ldots             		  \\
{He~II~$\lambda$4685}       	& 0.007$\pm$0.001  	& 0.021$\pm$0.001  	& 0.010$\pm$0.001  	& $<\sim$~0.009      	& 0.003$\pm$0.001  	& 0.006$\pm$0.002  	& 0.022$\pm$0.001    \\
{H$\beta$ $\lambda$4861}    	& 1.000$\pm$0.020  	& 1.000$\pm$0.020  	& 1.000$\pm$0.020  	& 1.000$\pm$0.020  	& 1.000$\pm$0.020  	& 1.000$\pm$0.020  	& 1.000$\pm$0.020    \\
{He~I~$\lambda$4921}        	& 0.007$\pm$0.001  	& 0.013$\pm$0.001  	& 0.008$\pm$0.001  	& $<\sim$~0.013      	& 0.008$\pm$0.001  	& 0.006$\pm$0.002  	& 0.010$\pm$0.001    \\
{[O~\iii]~$\lambda$4958}    	& 0.033$\pm$0.001  	& 0.169$\pm$0.003  	& 0.049$\pm$0.001  	& 0.056$\pm$0.004  	& 0.051$\pm$0.001  	& 0.112$\pm$0.002  	& 0.047$\pm$0.001    \\
{[O~\iii]~$\lambda$5006}    	& 0.101$\pm$0.002  	& 0.507$\pm$0.010  	& 0.144$\pm$0.003  	& 0.176$\pm$0.004  	& 0.155$\pm$0.003  	& 0.337$\pm$0.007  	& 0.154$\pm$0.003    \\
{He~I~$\lambda$5015}        	& 0.015$\pm$0.001  	& 0.020$\pm$0.001  	& 0.013$\pm$0.001  	& $<\sim$~0.012      	& 0.013$\pm$0.001  	& 0.009$\pm$0.002  	& 0.015$\pm$0.001    \\
{NI $\lambda$5197}          		& 0.009$\pm$0.001  	& 0.008$\pm$0.001  	& 0.009$\pm$0.001  	& $<\sim$~0.012      	& 0.015$\pm$0.001  	& 0.012$\pm$0.002  	& 0.003$\pm$0.001    \\
{O~I~$\lambda$5577}         	& 0.003$\pm$0.001  	& $<\sim$~0.004      	& $<\sim$~0.002      	& $<\sim$~0.015      	& $<\sim$~0.003      	& \ldots            		& 0.004$\pm$0.001    \\
{[N~\ii]~$\lambda$5754}     	& 0.004$\pm$0.001  	& 0.006$\pm$0.001  	& $<\sim$~0.001      	& 0.006$\pm$0.001  	& 0.005$\pm$0.001  	& 0.008$\pm$0.001  	& $<\sim$~0.001        \\
{He~I~$\lambda$5875}        	& 0.074$\pm$0.001  	& 0.104$\pm$0.002  	& 0.069$\pm$0.001  	& 0.078$\pm$0.002  	& 0.084$\pm$0.002  	& 0.088$\pm$0.002  	& 0.101$\pm$0.002    \\
{[O~\i] $\lambda$6300}      	& 0.015$\pm$0.001  	& 0.018$\pm$0.001  	& 0.019$\pm$0.001  	& 0.014$\pm$0.002  	& 0.020$\pm$0.001  	& 0.022$\pm$0.001  	& 0.008$\pm$0.001    \\
{[S~\iii]~$\lambda$6312}    	& 0.003$\pm$0.001  	& 0.007$\pm$0.001  	& 0.004$\pm$0.001  	& 0.004$\pm$0.001  	& $<\sim$~0.001      	& 0.007$\pm$0.001  	& 0.003$\pm$0.001    \\
{[O~\i] $\lambda$6363}      	& 0.004$\pm$0.001  	& 0.005$\pm$0.001  	& 0.007$\pm$0.001  & $<\sim$~0.005      	& 0.007$\pm$0.001  	& 0.007$\pm$0.001  	& 0.003$\pm$0.001    \\
{[N~\ii]~$\lambda$6548}     	& 0.321$\pm$0.006  	& 0.342$\pm$0.007  	& 0.307$\pm$0.006  & 0.316$\pm$0.006  	& 0.312$\pm$0.006  	& 0.309$\pm$0.006  	& 0.294$\pm$0.006    \\
{H$\alpha$ $\lambda$6562}   	& 3.095$\pm$0.062  	& 3.029$\pm$0.061  	& 3.103$\pm$0.062  	& 3.063$\pm$0.061  	& 3.105$\pm$0.062  	& 3.074$\pm$0.061  	& 3.182$\pm$0.064    \\
{[N~\ii]~$\lambda$6583}     	& 1.008$\pm$0.020  	& 1.055$\pm$0.021  	& 0.957$\pm$0.019  	& 0.987$\pm$0.020  	& 0.977$\pm$0.020  	& 1.025$\pm$0.020  	& 0.898$\pm$0.018    \\
{He~I~$\lambda$6678}        	& 0.025$\pm$0.001  	& 0.034$\pm$0.001  	& 0.022$\pm$0.001  	& 0.023$\pm$0.002  	& 0.023$\pm$0.001  	& 0.027$\pm$0.001  	& 0.031$\pm$0.001    \\
{[S~\ii]~$\lambda$6716}     	& 0.313$\pm$0.006  	& 0.234$\pm$0.005  	& 0.401$\pm$0.008  	& 0.288$\pm$0.006  	& 0.423$\pm$0.008  	& 0.333$\pm$0.007  	& 0.243$\pm$0.005    \\
{[S~\ii]~$\lambda$6730}     	& 0.225$\pm$0.005  	& 0.192$\pm$0.004  	& 0.289$\pm$0.006  	& 0.204$\pm$0.004  	& 0.306$\pm$0.006  	& 0.253$\pm$0.005  	& 0.170$\pm$0.003    \\
{He~I~$\lambda$7065}        	& 0.010$\pm$0.001  	& 0.020$\pm$0.001  	& 0.013$\pm$0.001  	& 0.009$\pm$0.001  	& 0.010$\pm$0.001  	& 0.017$\pm$0.001  	& 0.014$\pm$0.001    \\
{[Ar~\iii]~$\lambda$7135}   	& 0.028$\pm$0.001  	& 0.069$\pm$0.001  	& 0.026$\pm$0.001  	& 0.034$\pm$0.001  	& 0.025$\pm$0.001  	& 0.057$\pm$0.001  	& 0.036$\pm$0.001    \\
{He~I~$\lambda$7281}        	& 0.004$\pm$0.001  	& 0.005$\pm$0.001  	& 0.003$\pm$0.001  	& $<\sim$~0.004      	& 0.003$\pm$0.001  	& 0.004$\pm$0.001  	& 0.004$\pm$0.001    \\
{[O~\ii]~$\lambda$7319}     	& $<\sim$~0.004      	& 0.013$\pm$0.001  	& 0.007$\pm$0.001  	& 0.007$\pm$0.001  	& 0.008$\pm$0.001  	& 0.011$\pm$0.001  	& $<\sim$~0.005        \\
{[O~\ii]~$\lambda$7330}     	& 0.004$\pm$0.001  	& 0.010$\pm$0.001  	& 0.003$\pm$0.001  	& 0.004$\pm$0.001  	& 0.007$\pm$0.001  	& 0.011$\pm$0.001  	& $<\sim$~0.004        \\
{[Ar~\iii]~$\lambda$7751}   	& 0.008$\pm$0.001  	& 0.017$\pm$0.001  	& 0.008$\pm$0.001  	& 0.008$\pm$0.001  	& 0.008$\pm$0.001  	& 0.013$\pm$0.001  	& 0.011$\pm$0.001    \\
{P18 $\lambda$8437}         	& 0.004$\pm$0.001  	& 0.004$\pm$0.001  	& 0.004$\pm$0.001  	& $<\sim$~0.019      	& 0.006$\pm$0.001  	& 0.011$\pm$0.002  	& 0.007$\pm$0.001    \\
{O~I~$\lambda$8446}         	& 0.006$\pm$0.001  	& 0.007$\pm$0.001  	& 0.011$\pm$0.001  	& $<\sim$~0.020      	& 0.007$\pm$0.001  	& \ldots            		& \ldots             		  \\
{P17 $\lambda$8467}         	& 0.004$\pm$0.001  	& 0.005$\pm$0.001  	& 0.008$\pm$0.001  	& $<\sim$~0.019      	& 0.004$\pm$0.001  	& $<\sim$~0.005      	& 0.006$\pm$0.001    \\
{P16 $\lambda$8502}         	& 0.006$\pm$0.001  	& 0.006$\pm$0.001  & 0.009$\pm$0.001  	& $<\sim$~0.020      	& 0.005$\pm$0.001  	& $<\sim$~0.005      	& 0.005$\pm$0.001    \\
{P15 $\lambda$8545}         	& 0.007$\pm$0.001  	& 0.007$\pm$0.001  	& 0.012$\pm$0.001  	& $<\sim$~0.019      	& 0.006$\pm$0.001  	& $<\sim$~0.005      	& 0.007$\pm$0.001    \\
{P14 $\lambda$8598}         	& 0.007$\pm$0.001  	& 0.008$\pm$0.001  	& 0.013$\pm$0.001  	& $<\sim$~0.020      	& 0.007$\pm$0.001  	& 0.008$\pm$0.002  	& 0.008$\pm$0.001    \\
{P13 $\lambda$8665}         	& 0.009$\pm$0.001  	& 0.011$\pm$0.001  	& 0.017$\pm$0.001  	& $<\sim$~0.021      	& 0.009$\pm$0.001  	& 0.008$\pm$0.002  	& 0.010$\pm$0.001    \\
{N~I~$\lambda$8683}         	& 0.002$\pm$0.001  	& $<\sim$~0.004      	& 0.006$\pm$0.001  	& $<\sim$~0.021      	& $<\sim$~0.002      	& $<\sim$~0.005      	& $<\sim$~0.002        \\
{P12 $\lambda$8750}         	& 0.005$\pm$0.001  	& 0.013$\pm$0.001  	& 0.012$\pm$0.001  	& $<\sim$~0.021      	& 0.020$\pm$0.001  	& 0.037$\pm$0.002  	& 0.012$\pm$0.001    \\
{P11 $\lambda$8862}         	& 0.010$\pm$0.001  	& 0.019$\pm$0.001  	& 0.021$\pm$0.001  	& $<\sim$~0.022      	& 0.017$\pm$0.001  	& 0.028$\pm$0.002  	& 0.023$\pm$0.001    \\
{P10 $\lambda$9015}         	& 0.024$\pm$0.001  	& 0.022$\pm$0.001  	& 0.024$\pm$0.001  	& 0.021$\pm$0.002  	& 0.024$\pm$0.001  	& 0.017$\pm$0.001  	& 0.025$\pm$0.001    \\
{[S~\iii]~$\lambda$9068}    	& 0.238$\pm$0.005  	& 0.367$\pm$0.007  	& 0.192$\pm$0.004  	& 0.245$\pm$0.005  	& 0.167$\pm$0.003  	& 0.293$\pm$0.006  	& 0.284$\pm$0.006    \\
{P9 $\lambda$9229}          	& 0.030$\pm$0.001  	& 0.030$\pm$0.001  	& 0.032$\pm$0.001  	& 0.031$\pm$0.003  	& 0.025$\pm$0.001  	& 0.029$\pm$0.001  	& 0.036$\pm$0.001    \\
{[S~\iii]~$\lambda$9530}    	& 0.644$\pm$0.013  	& 0.969$\pm$0.019  	& 0.554$\pm$0.011  	& 0.699$\pm$0.015  	& 0.479$\pm$0.010  	& 0.837$\pm$0.017  	& 0.788$\pm$0.016    \\
{P8 $\lambda$9546}          	& 0.020$\pm$0.001  	& 0.044$\pm$0.001  	& 0.025$\pm$0.001  	& 0.019$\pm$0.005  	& 0.054$\pm$0.001  	& 0.030$\pm$0.001  	& 0.061$\pm$0.001    \\
\hline
{C$_{H\beta}$}              		& 0.297$\pm$0.013  	& 0.450$\pm$0.013 	& 0.221$\pm$0.013  	& 0.250$\pm$0.013  	& 0.433$\pm$0.013  	& 0.542$\pm$0.013 	& 0.148$\pm$0.013  	\\
{F$_{H\beta}$}              		& 257.9$\pm$0.4     	& 715.5$\pm$2.9     	& 391.8$\pm$0.3     	& 202.3$\pm$1.2     	& 190.2$\pm$0.3     	& 77.9$\pm$0.2      	& 297.6$\pm$0.2      	\\
{EW$_{H\beta}$}             		& 202.3             	& 139.8             	& 102.6             	& 143.3             	& 121.6             	& 90.9              		& 114.5              	\\
{EW$_{H\alpha}$}            		& 1331.5            	& 1109.7            	& 647.0             	& 826.5             	& 910.7             	& 666.0             	& 863.8              	\\
\enddata
\tablecomments{\ Emission line fluxes. Detections at a significance of less than 3 $\sigma$ are given as upper limits.
Full table available online only.}
\label{tbl3}
\end{deluxetable*}


\section{Physical Conditions}\label{sec:phys}

\subsection{Three-Zone Temperature Determinations}\label{sec:zones}

A simple \ion{H}{2} region can be modeled by three separate volumes: 
a low-, intermediate-, and high-ionization zone.
Accurate \ion{H}{2} region abundance determinations require reliable electron temperature measurements for each volume.
This is typically done by observing a temperature-sensitive auroral-to-strong-line ratio. 
The [\ion{O}{3}] I($\lambda\lambda$4959,5007)/I($\lambda$4363) ratio 
is expected to reflect the temperature in the high ionization zone.
Similarly, the [\ion{N}{2}] I($\lambda\lambda$6548,6484)/I($\lambda$5755) and 
the [\ion{O}{2}] I($\lambda\lambda$7320,7330)/I($\lambda$3726,3728) ratios
are expected to reflect the temperature in the low ionization zone.
S and Ar ions do not neatly separate into these two zones and thus 
require an intermediate ionization zone, which can be measured 
by the [\ion{S}{3}] I($\lambda\lambda$9069,9532)/I($\lambda$6312) ratio.
Since the redder [\ion{S}{3}] lines can be affected by atmospheric absorption,
we employ the theoretical ratio of $\lambda$9532/$\lambda$9069 = 2.51 to 
make an upward correction to the weaker of the two lines if a deviation
of more than 2\% from the theoretical ratio is observed.

Following the precedent of \citet{bresolin09a}, we have adopted the current best 
set of atomic data from the literature as reported in Table~\ref{tbl4}. 
For the \ion{H}{2} regions with detected auroral lines, we determine electron temperatures using the reddening 
corrected line-ratios and appropriate atomic data, assuming the ions are well-approximated by a 5-level atom\footnotemark[14].
Once a direct electron temperature is determined, 
the physical conditions of the other zones are needed to complete the \ion{H}{2} region picture.
\citet[hereafter G92;][]{garnett92} relates the direct temperatures of different ionization zones using photoionization models: 
\begin{align}
       \mbox{T[S~\iii]} & =  0.83\times \mbox{T[O~\iii]} + 1700\mbox{ K} \label{eqn:G92-1} \\
       \mbox{T[N~\ii]} & =  0.70\times \mbox{T[O~\iii]} + 3000\mbox{ K.} \label{eqn:G92-2}
\end{align}
These relationships are valid for temperatures typical of \ion{H}{2} regions: $T_e$ = 2,000$-$18,000\,K.
Observational evidence supporting these relationships can be found in the literature.
For example, \citet{bresolin09a} and \citet{kennicutt03a} find a tight correlation 
between \ion{H}{2} region observations and the G92 T[\ion{S}{3}] versus 
T[\ion{O}{3}] relationship and \citet{esteban09} find a high 
correlation coefficient for the T[\ion{N}{2}] versus T[\ion{O}{3}] G92 model.  
With the assumption that T[\ion{N}{2}]  = T[\ion{O}{2}], 
since N$^{+}$ and O$^{+}$ have comparable ionization energies, 
we now have the tools to determine temperatures in all three ionization zone for all ions.

\footnotetext[14]{https://github.com/moustakas/impro} 


\subsection{Impact Of Updated Atomic Data}\label{sec:atomic}

Recent updates to atomic data are of particular importance to this paper
as changes in the Einstein A-coefficients for [\ion{S}{3}] result in 
a systematic shift of T[\ion{S}{3}] to cooler temperatures.
In M101, \citet[][hereafter K03]{kennicutt03a} 
measured 16 regions with both T[\ion{O}{3}] and T[\ion{S}{3}] 
and found a tight correlation around relationship defined by G92.
K03 used collision strengths for S$^{++}$ from \citet{tayal99} 
and Einstein A coefficients from \citet{kauffman86}. 
Contemporaneously, \citet{perez-montero03} noted a T[\ion{S}{3}]-T[\ion{O}{3}] 
theoretical relationship that differs from the standard G92 relationship, where 
the departure is mostly due to introducing atomic data from \citet{tayal99}.
This is also a factor in the analysis by \citet{binette12}, who use the 
abundances from \citet{hagele06} and \citet{perez-montero06}, both of 
whom employ the [\ion{S}{3}] collision strengths from \citet{tayal99}. 
Since then, further derivations of atomic coefficients for S$^{++}$ have been published, 
such as the work of \citet{ff06}, whose Einstein A-values we adopt.
\citet{mendoza14} performed extensive computations of new collisional data sets for
[\ion{O}{3}], [\ion{O}{2}], [\ion{N}{2}], [\ion{S}{3}], [\ion{S}{2}], and [\ion{Ar}{3}] and compared
to other data sets of effective collision strengths computed within the past two decades.
They found that statistical consistency amongst data sets is typically around $\sim20-30\%$,
where the inherent dispersion in forbidden transitions is due to the strong sensitivity of the the 
resonance contribution to the scattering numerical approach, target atomic model, energy mesh, 
and small interactions such as relativistic corrections. 
However, they note that with the exception of [\ion{S}{2}], the resulting theoretical temperature 
dispersions have been found to be no larger than 10$\%$.
Following the recommendations of \citet{mendoza14}, and considering the discussion of
effective collision strengths for [\ion{O}{3}] by \citet{storey14},
we adopt the most up-to-date atomic data (see Table~\ref{tbl4}).

\begin{figure}
\epsscale{1.2}
\plotone{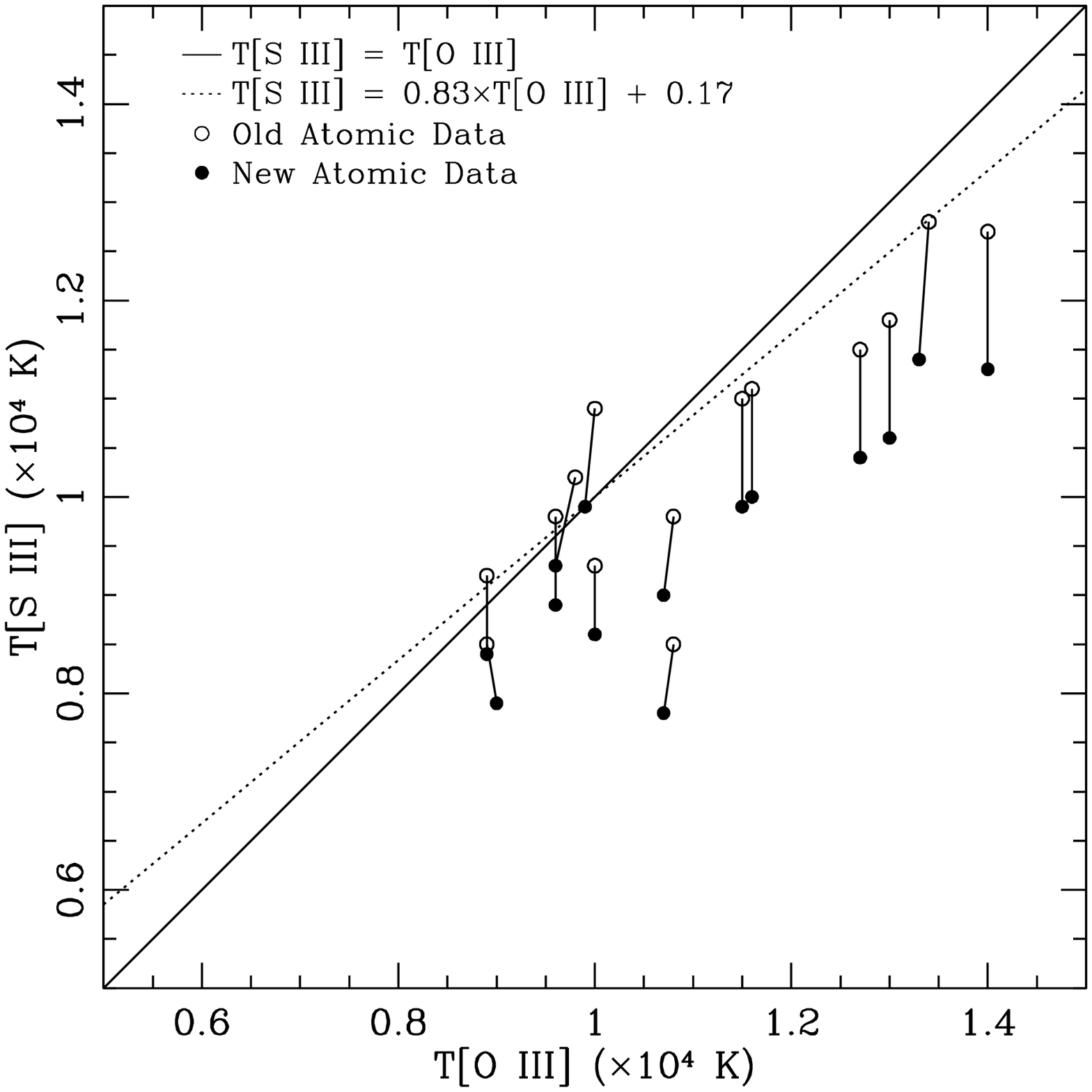}
\caption{The effect of updating atomic data as demonstrated by comparisons of
[\ion{O}{3}] and [\ion{S}{3}] temperature determinations for the \ion{H}{2} regions in M101 from K03.
Temperatures determined with the atomic dataset of K03 (specifically, \citet{tayal99} and 
\citet{kauffman86} for [\ion{S}{3}]) are plotted as open circles, whereas values determined 
using the sources adopted here (see Table~\ref{tbl4}) are closed circles.
Sets of points belonging to the same \ion{H}{2} region are connected.
For reference, the equivalence relation is plotted as a solid line and the theoretical 
relationship of G92 is shown as a short dashed line.}
\label{fig3}
\end{figure}

\begin{figure}
\epsscale{1.2}
\plotone{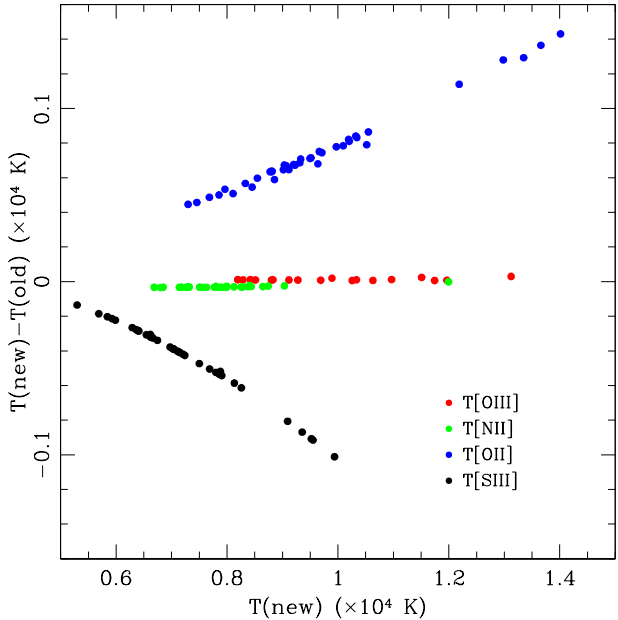}
\caption{The effect of updating atomic data as demonstrated by 
the difference of new and old electron temperatures versus new temperature.
Data is color coded for each ion: T[\ion{O}{3}] - red, T[\ion{N}{2}] - orange, 
[\ion{O}{2}] - green, and T[\ion{S}{3}] - blue. 
Little difference is seen for T[\ion{N}{2}] and T[\ion{O}{3}], 
whereas significant updates have been made to the [\ion{S}{3}] and [\ion{O}{2}] atomic data
and resulting temperatures.}
\label{fig4}
\end{figure}

To demonstrate the significance of atomic data updates, in Figure~\ref{fig3} we present a 
comparison of temperatures determined for the K03 M101 regions using 
(1) the atomic data as in KO3: specifically, \citet{tayal99} and \citet{kauffman86} for [\ion{S}{3}], and 
(2) the current database adopted here (see Table~\ref{tbl4}).
Temperatures determined with the old atomic data are plotted as open circles, 
while the updated values are closed circles, 
and the set of points belonging to the same \ion{H}{2} region are connected.
For reference, the one-to-one relation is plotted as a solid line and the 
theoretical relationship of G92 is shown as a short dashed line.
The comparison in Figure~\ref{fig3} shows a shift of roughly 10\% in T[\ion{S}{3}] ($\sim$1000 K) in
the sense that using the newer atomic data results in lower values of T[\ion{S}{3}].
Additionally, in Figure~\ref{fig4} we plot the difference in electron temperature due to updating 
the atomic data versus the new electron temperature.
The old temperatures are determined using the same atomic data as K03, whereas the new 
temperatures use the updated atomic data listed in Table~\ref{tbl4}.
The temperature trends shown in Figure~\ref{fig4} show that the atomic data for [\ion{N}{2}]
and [\ion{O}{3}] have remained fairly stable, whereas the data for [\ion{O}{2}] and [\ion{S}{3}]
have not yet converged.
As we will see in the next section, our observations show significant discrepancies between values of
T[\ion{S}{3}] and T[\ion{O}{3}], in the sense that some T[\ion{O}{3}] values are significantly higher.


\subsection{Electron Temperature and Density Determinations}\label{sec:temden}

To establish a low-uncertainty electron temperature estimate, we define a quality criterion 
requiring a 3$\sigma$ detection of the auroral line measurement (see \S~\ref{sec:iraf}).
The high throughput of the MODS over a large optical and near-UV wavelength range
allows us to take advantage of the following auroral lines: 
[\ion{S}{2}] $\lambda\lambda$4068,4076,
[\ion{O}{3}] $\lambda$4363, [\ion{Ar}{3}] $\lambda$5192, [\ion{N}{2}] $\lambda$5755, 
[\ion{S}{3}] $\lambda$6312, and [\ion{O}{2}] $\lambda\lambda$7320,7330.
In the NGC~628 spectra, there are very few [\ion{S}{2}] $\lambda\lambda$4068,4076 and
[\ion{Ar}{3}] $\lambda$5192 detections, so we will focus on the four lines with the most numerous detections. 
We detect the following auroral line measurements at a strength of 3$\sigma$ or greater:
[\ion{O}{3}] $\lambda$4363 in 18 \ion{H}{2} regions,  
[\ion{N}{2}] $\lambda$5755 in 29 \ion{H}{2} regions,
[\ion{S}{3}] $\lambda$6312 in 40 \ion{H}{2} regions, and
[\ion{O}{2}] $\lambda\lambda$7320,7330 in 40 \ion{H}{2} regions.
Due to the previous concerns with [\ion{O}{2}] temperature measurements \citep[e.g.,][]{kennicutt03a}, 
we do not report abundances calculated from this diagnostic. 
Rather, we defer to a future analysis of the entire CHAOS sample.
The result is a sample of 45 \ion{H}{2} regions with direct abundances from [\ion{O}{3}], 
[\ion{N}{2}], or [\ion{S}{3}]; where 40 of these regions have multiple temperature determinations. 
The auroral lines measured for each \ion{H}{2} region are listed in Table~\ref{tbl2}.

The [\ion{S}{2}] $\lambda\lambda$6717,6731 ratio is used to determine the electron densities, which are 
confirmed with density determinations from the [\ion{O}{2}] $\lambda\lambda$3726,3728 line ratio.\footnotemark[15]
For all densities which are consistent with the low density limit, abundance calculations assume $n_e = 10^2$ cm$^{-3}$ 
(which is consistent with the 1$\sigma$ upper bounds and produces identical results for all lower values of $n_e$),
else the derived higher electron density ($n_e > 10^2$ cm$^{-3}$) is used. 
The electron temperature and density determinations are listed in Table~\ref{tbl5} (full table is available online).

\footnotetext[15]{At the resolution of MODS, the [\ion{O}{2}] $\lambda\lambda3726,3728$ 
doublet is partially blended for all observations, and so measuring this ratio serves primarily 
as a sanity check of the [\ion{S}{2}] density determination.}


\begin{deluxetable*}{ccc}
\tabletypesize{\scriptsize}
\setlength{\tabcolsep}{3pt} 
\tablewidth{0pt}
\tablecaption{Sources of Updated Atomic Data for CHAOS}
\tablehead{
Ion			& Radiative Transition Probabilities	& Collision Strengths }
\startdata
{[\ion{O}{2}]}	& {\citet{fft04}$^{\star\dagger}$}	& {\citet{kisielius09}}	\\
{[\ion{O}{3}]}	& {\citet{fft04}$^{\star\dagger}$}	& {\citet{storey14}}		\\
{[\ion{N}{2}]}	& {\citet{fft04}$^{\star}$}		& {\citet{tayal11}}		\\
{[\ion{Ne}{3}]}	& {\citet{fft04}$^{\star}$}		& {\citet{mclaughlin00}} 	\\
{[\ion{Ar}{3}]}	& {\citet{mendoza83}}		& {\citet{munoz-burgos09}} \\
{[\ion{S}{2}]}	& {\citet{mendoza14}}		& {\citet{tayal10}} 		\\
{[\ion{S}{3}]}	& {\citet{ff06}}				& {\citet{hudson12}}		 \\
\enddata
\tablecomments{
The atomic data used by the 5-level atom approximation has been updated with
current values from the literature as reported here. \\
$^{\star}$ Agrees with updated values from \citet{tayal11}. \\
$^{\dagger}$ Equivalent to \citet{tachiev02}, as recommended by \citet{stasinska12}.}
\label{tbl4}
\end{deluxetable*}		

\begin{deluxetable*}{lcccccc}
\tablewidth{0pt}	
\tabletypesize{\scriptsize}
\setlength{\tabcolsep}{3pt} 
\tablecaption{Ionic and Total Abundances for MODS/LBT Observations of NGC 628}
\tablehead{   	
\CH{H$\alpha$ Region} 	        		& \CH{-35.9+57.7}    	& \CH +49.8+48.7    	& \CH{-73.1-27.3}    	& \CH{-76.2+22.9 }  	& \CH{-36.8-73.4}   	& \CH{+68.5+53.4}  		}

t(O$_2$)$_{measured}$ (K)        	& \ldots            		& 9000$\pm$100      	& 7300$\pm$100      	& 7500$\pm$500      	& 8200$\pm$100      	& 9300$\pm$200      	\\
t(O$_3$)$_{measured}$ (K)        	& \ldots            		& \ldots            		& \ldots            		& \ldots            		& \ldots            		& \ldots                     	\\
t(N$_2$)$_{measured}$ (K)        	& 6800$\pm$100      & 7200$\pm$200      	& \ldots            		& 7300$\pm$400      	& 7300$\pm$200      	& 8100$\pm$300      	\\
t(S$_3$)$_{measured}$ (K)        	& 5700$\pm$100      & 6400$\pm$100      	& 6400$\pm$100      	& 6000$\pm$400      	& \ldots            		& 6600$\pm$200      	\\
$n_e$($_{measured}$) (cm$^{-3}$) 	& 57                		& 248               		& 59                		& 37                		& 64                		& 126                      	\\
\vspace{-0.10cm} \\
t(O$_2$)$_{used}$ (K)            		& 6800$\pm$500      & 7200$\pm$200      	& 6900$\pm$500      	& 7300$\pm$400      	& 7300$\pm$200      	& 8100$\pm$300      	\\
t(O$_3$)$_{used}$ (K)            		& 4800$\pm$500      & 5700$\pm$500      	& 5600$\pm$500      	& 5200$\pm$500      	& 6100$\pm$500      	& 6000$\pm$500      	\\
t(N$_2$)$_{used}$ (K)            		& 6800$\pm$200      & 7200$\pm$200      	& 6900$\pm$500      	& 7300$\pm$400      	& 7300$\pm$200      	& 8100$\pm$300      	\\
t(S$_3$)$_{used}$ (K)            		& 5700$\pm$200      & 6400$\pm$200      	& 6400$\pm$200      	& 6000$\pm$400      	& 6800$\pm$500      	& 6600$\pm$200      	\\
$n_e$($_{used}$) (cm$^{-3}$)     	& 100               		& 248               		& 100               		& 100               		& 100               		& 126                        	\\
\hline
O+/H+ (10$^{5}$)                 		& 28.1$\pm$1.4      	& 32.5$\pm$2.0      	& 26.9$\pm$5.3      	& 20.4$\pm$3.2      	& 21.4$\pm$1.3      	& 14.2$\pm$1.2        	\\
O++/H+ (10$^{5}$)                		& 11.7$\pm$3.8      	& 22.1$\pm$5.2      	& 6.58$\pm$1.55     	& 12.9$\pm$3.9      	& 4.44$\pm$0.90     	& 11.3$\pm$2.4        	\\
12 + log(O/H) (dex)              		& 8.60$\pm$0.04     	& 8.74$\pm$0.04     	& 8.52$\pm$0.07     	& 8.52$\pm$0.07     	& 8.41$\pm$0.03     	& 8.41$\pm$0.05      	\\
\hline
N+/H+ (10$^{6}$)                 		& 64.8$\pm$2.1      	& 56.5$\pm$2.2      	& 57.2$\pm$6.9      	& 49.1$\pm$4.8      	& 49.2$\pm$1.8      	& 35.2$\pm$1.8        	\\
N ICF                            			& 1.417$\pm$0.161  	& 1.680$\pm$0.200  	& 1.245$\pm$0.318 	& 1.635$\pm$0.361  	& 1.207$\pm$0.101  	& 1.796$\pm$0.240  	\\
log(N/O) (dex)                   			& -0.64$\pm$0.03    	& -0.76$\pm$0.03    	& -0.67$\pm$0.10    	& -0.62$\pm$0.08    	& -0.64$\pm$0.03    	& -0.61$\pm$0.04     	\\
12 + log(N/H) (dex)              		& 7.96$\pm$0.05     	& 7.98$\pm$0.05     	& 7.85$\pm$0.12     	& 7.90$\pm$0.10     	& 7.77$\pm$0.04     	& 7.80$\pm$0.06      	\\
\hline
S+/H+ (10$^{7}$)                 		& 43.1$\pm$1.4      	& 29.2$\pm$1.1      	& 51.4$\pm$6.0      	& 30.7$\pm$2.9      	& 46.1$\pm$1.6      	& 26.7$\pm$1.3        	\\
S++/H+ (10$^{7}$)                		& 178.$\pm$5.4      	& 188.$\pm$4.4      	& 104.$\pm$2.2      	& 160.$\pm$17.      	& 75.6$\pm$7.4      	& 141.$\pm$5.3        	\\
S ICF                            			& 1.083$\pm$0.108  	& 1.151$\pm$0.115 	& 1.026$\pm$0.103  	& 1.142$\pm$0.114  	& 1.013$\pm$0.101  	& 1.172$\pm$0.117  	\\
log(S/O) (dex)                   			& -1.22$\pm$0.06    	& -1.34$\pm$0.06    	& -1.32$\pm$0.09    	& -1.19$\pm$0.09    	& -1.32$\pm$0.06    	& -1.11$\pm$0.06     	\\
12 + log(S/H) (dex)              		& 7.38$\pm$0.05     	& 7.40$\pm$0.04     	& 7.20$\pm$0.05     	& 7.34$\pm$0.06     	& 7.09$\pm$0.05     	& 7.29$\pm$0.05      	\\
\hline
Ne++/H+ (10$^{6}$)               		& \ldots            		& 24.8$\pm$0.58     	& 14.1$\pm$2.8      	& \ldots            		& \ldots            		& 26.9$\pm$0.63      	\\
Ne ICF                           			& 3.398$\pm$1.144  	& 2.471$\pm$0.628  	& 5.085$\pm$1.461  	& 2.575$\pm$0.878 	& 5.823$\pm$1.224  	& 2.256$\pm$0.535  	\\
log(Ne/O) (dex)                  		& \ldots            		& -0.95$\pm$0.10    	& -0.67$\pm$0.13    	& \ldots            		& \ldots            		& -0.62$\pm$0.09     	\\
12 + log(Ne/H) (dex)             		& \ldots            		& 7.79$\pm$0.11     	& 7.85$\pm$0.15     	& \ldots            		& \ldots            		& 7.78$\pm$0.10      	\\
\hline
Ar++/H+ (10$^{7}$)               		& 14.7$\pm$0.53     	& 21.5$\pm$0.56     	& 9.08$\pm$0.22     	& 13.3$\pm$2.0      	& 7.02$\pm$0.83     	& 14.9$\pm$0.69      	\\
Ar ICF                           			& 1.916$\pm$0.192  	& 1.569$\pm$0.157  	& 2.730$\pm$0.273  	& 1.602$\pm$0.160  	& 3.140$\pm$0.314 	& 1.510$\pm$0.151  	\\
log(Ar/O) (dex)                  			& -2.15$\pm$0.06    	& -2.21$\pm$0.06    	& -2.13$\pm$0.08    	& -2.19$\pm$0.10    	& -2.07$\pm$0.07    	& -2.05$\pm$0.07     	\\
12 + log(Ar/H) (dex)             		& 6.45$\pm$0.05     	& 6.53$\pm$0.05     	& 6.39$\pm$0.05     	& 6.33$\pm$0.08     	& 6.34$\pm$0.07     	& 6.35$\pm$0.05   

\enddata 
\tablecomments{Electron temperatures and ionic and total abundances for objects with an 
[\ion{O}{3}] $\lambda4363$, [\ion{N}{2}] $\lambda5755$, or  [\ion{S}{3}] $\lambda6312$ 
line signal to noise ratio of $3\sigma$ or greater. Electron temperatures were calculated using the 
[\ion{O}{3}] ($\lambda4959 + \lambda5007$)/$\lambda4363$, [\ion{N}{2}] 
($\lambda6548 + \lambda6584$)/$\lambda5755$, or the [\ion{S}{3}] 
($\lambda9069 + \lambda9532$)/$\lambda6312$  diagnostic line ratio.
Full table available online only.} 
\label{tbl5}
\end{deluxetable*}


\begin{figure*}
\begin{tabular}{cc}
	\includegraphics[scale = 1.1, trim = 10mm 5mm 120mm 0, clip]{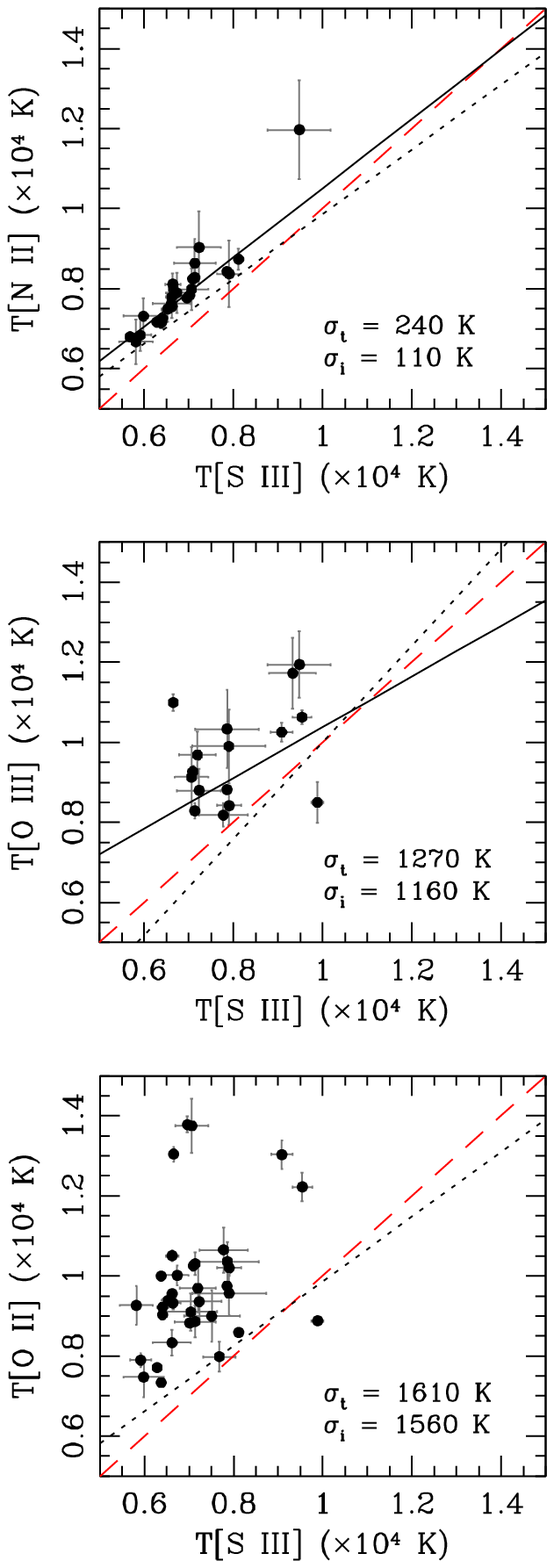} & \includegraphics[scale = 1.1, trim = 120mm 5mm 0 0, clip]{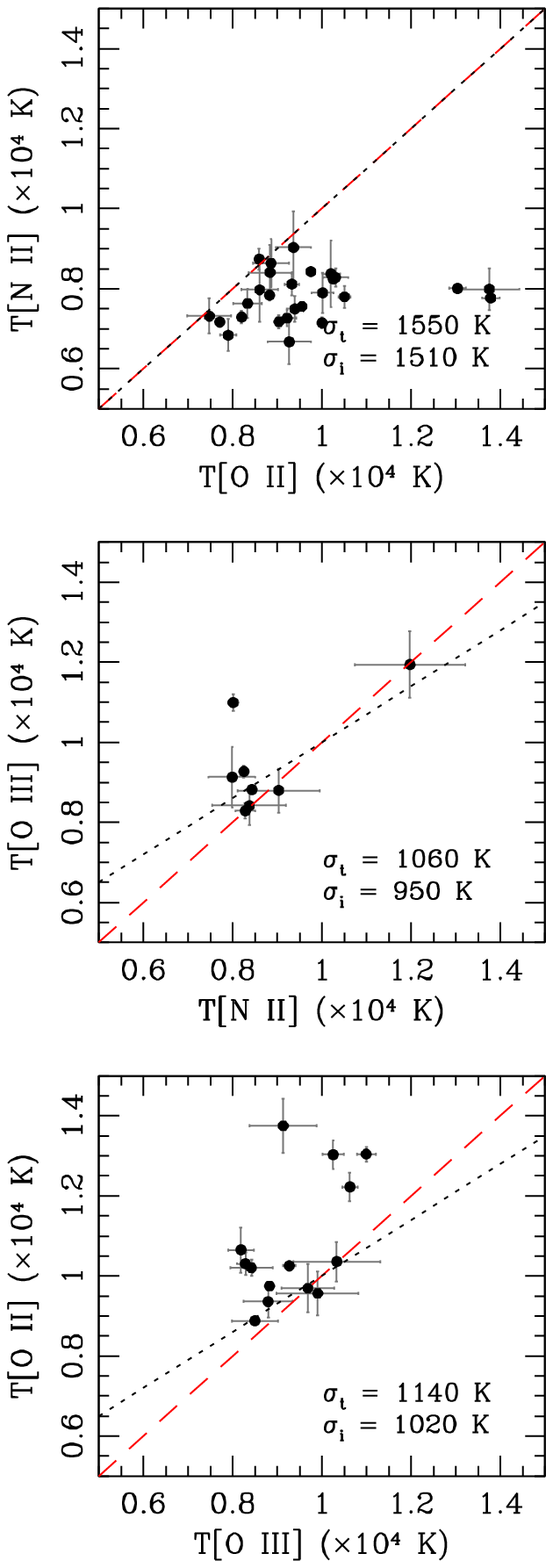}
\end{tabular}
\caption{Comparisons of the relations between the [\ion{O}{3}], [\ion{N}{2}], [\ion{O}{2}],
and [\ion{S}{3}] $T_e$ temperatures for \ion{H}{2} regions with more than one
temperature measurement.
The red long-dashed lines represent equal temperatures and black short-dashed lines are the G92 relationships.
For the T[\ion{N}{2}] versus T[\ion{S}{3}] and T[\ion{O}{3}] versus T[\ion{S}{3}] panels, 
the error-weighted least-squares best fit is shown by solid lines.
The total and intrinsic scatter around the line of equality is give for each comparison.}
\label{fig5}
\end{figure*}


\subsection{Comparing Electron Temperature Measurements}\label{sec:Tcomp}

Using the 45 \ion{H}{2} regions in our sample with multiple auroral line measurements,
all of which were double checked by hand,
we compare the calculated electron temperatures in Figure~\ref{fig5}.
The red dashed lines show the line of equality and the black short dashed lines are the expected 
relationships from the photoionization models calculated by G92.
For each set of variables we calculate the scatter in the data, with errors
in both coordinates, around the line of equality and determine how much can 
be attributed to a physically significant intrinsic scatter using the MPFITEXY 
\citep{williams10}\footnotemark[16] routine in IDL.
The intrinsic scatter is determined by minimizing the departure of $\chi^2$/(degrees of freedom) 
from 1 by introducing an additional scatter term to the weighting of each data point. 
Additionally, in two cases we also perform a least-squares fit, which is shown with a solid line.
The calculated total and intrinsic scatters, $\sigma_t$ and $\sigma_i$ respectively, are displayed in Figure~\ref{fig5}.

\footnotetext[16]{MPFITEXY is dependent on the MPFIT package \citep{markwardt09}.}

The left column of Figure~\ref{fig5} shows T[\ion{N}{2}],
T[\ion{O}{3}], and T[\ion{O}{2}] as a function of T[\ion{S}{3}].
The top panel of this column shows a very tight correlation between T[\ion{N}{2}] and T[\ion{S}{3}].  
The dispersion in this relationship about the line of equality is 240\,K, and is consistent 
with being due entirely to observational uncertainties (i.e., very little internal scatter). 
The middle panel shows a large scatter for the comparison of T[\ion{O}{3}] and T[\ion{S}{3}],
with a dispersion around the line of equality of 1270\,K, indicating an intrinsic dispersion of 1160\,K.
The bottom panel shows a comparison of the T[\ion{O}{2}] and T[\ion{S}{3}] measurements,
where the scatter is, again, quite large at 1610\,K, indicating an intrinsic dispersion of 1560\,K.
This panel is similar to the T[\ion{O}{3}] versus T[\ion{S}{3}] comparison 
in having both a large dispersion and a systematic offset to smaller T[\ion{S}{3}] values.

The right column of Figure~\ref{fig5} compares T[\ion{N}{2}],
T[\ion{O}{3}], and T[\ion{O}{2}] to one another.
The top panel shows T[\ion{N}{2}] versus T[\ion{O}{2}].  
In this comparison, the scatter is again large, with a dispersion of 1550\,K, indicating
that essentially all of the dispersion is intrinsic (formally 1510\,K).
In this comparison, where roughly equal temperatures are expected 
from photoionization models, the data appear to fall into two groups. 
Roughly half of the data are shifted from the equality relationship by
$\sim1400$\,K toward higher T[\ion{O}{2}].
The other half of the data show an even larger offset in the sense that the 
T[\ion{O}{2}] are hotter than the T[\ion{N}{2}] estimates.  
Of the outliers, three fall into a group where the average offset is close to 4000\,K.  
Most of the very large dispersion is due to this triad of outlying points.  
Overall, T[\ion{O}{2}] is systematically larger than both T[\ion{N}{2}] and T[\ion{S}{3}] 
as noted in previous studies \citep[e.g.,][]{pilyugin09,esteban09}.
The middle panel shows T[\ion{O}{3}] versus T[\ion{N}{2}].
Although there are only eight points in this comparison,  
all but one of the measurements are in agreement with the expected relationship from G92.  
Finally, in the bottom panel, the comparison between T[\ion{O}{2}] and T[\ion{O}{3}] shows
a large dispersion of 1140\,K, of which 1020\,K is due to intrinsic scatter.  
These points show a similar trend to that seen in the T[\ion{O}{2}]$-$T[\ion{S}{3}]
comparison, with preferential scatter to higher values of T[\ion{O}{2}].

The magnitude of the scatter in the T[\ion{O}{3}]$-$T[\ion{S}{3}] relationship came as a surprise.  
In fact, the CHAOS program was designed around the assumption that a secure 
measurement of T[\ion{O}{3}] implied a secure abundance (with the understanding 
that there could be small biases due to temperature inhomogeneities).
\citet{hagele06} and \citet{perez-montero06} have both shown a poor correlation
between T[\ion{O}{3}] and T[\ion{S}{3}], and \citet{binette12} showed definitively
that discrepancies between T[\ion{O}{3}] and T[\ion{S}{3}] are very common
in the sense that T[\ion{O}{3}] is higher than T[\ion{S}{3}]. 
We showed in Section~\ref{sec:atomic} that recent updates to atomic data enhances the
T[\ion{O}{3}]-T[\ion{S}{3}] discrepancy by uniformly shifting [\ion{S}{3}] temperatures cooler,
but cannot be entirely responsible for the magnitude of discrepancies observed.
Although many of the points in NGC~628 \textit{are} consistent with the equality trend 
or the relationship derived by G92, there are a significant number of outliers,  
in agreement with the trend noted by \citet{binette12}.  
This large number of outliers, and the similar trends noted in the literature, call into question 
the reliability of [\ion{O}{3}] as a temperature indicator in \ion{H}{2} regions.
We discuss the discrepant temperature measurements further in Section~\ref{sec:discrepant}.

In sum, the T[\ion{N}{2}] and T[\ion{S}{3}] measurements appear to be generally consistent, 
especially for T$_{e}<10^4$ K, while the T[\ion{O}{3}] and T[\ion{O}{2}] measurements are often discrepant.  
Because a significant fraction of the T[\ion{O}{3}] and T[\ion{O}{2}] measurements 
appear to be biased to higher than expected values, abundances based on the these 
temperature measurements exhibit both larger dispersions and systematic offsets.
These results guide our chosen methodology for deriving abundances in Section~\ref{sec:method}.


\section{A Homogeneous Abundance Analysis Based on [\ion{S}{3}] and [\ion{N}{2}] Temperatures}
\label{sec:abund}

\subsection{Methodology for Ionic and Absolute Abundance Calculations}\label{sec:method}

From the comparisons in Section~\ref{sec:Tcomp}, 
it is clear that there are significant temperature measurement discrepancies.  
Because T[\ion{N}{2}] and T[\ion{S}{3}] show a very good correlation at low temperatures,
we interpret this to mean that these temperatures are free from whatever systematic effects are
affecting the T[\ion{O}{3}] and T[\ion{O}{2}] measurements.
Fortunately, the vast majority of our sample have measurements of T[\ion{N}{2}] or T[\ion{S}{3}] (or both).
Thus, we calculate absolute and relative abundances based primarily on temperatures 
determined from T[\ion{N}{2}] and/or T[\ion{S}{3}] in conjunction with the G92
scaling relationships for the other ionization zones.  
Specifically, when T[\ion{N}{2}] and T[\ion{S}{3}] are both measured, T[\ion{N}{2}] is 
used for the low ionization zone,  T[\ion{S}{3}] is used for the intermediate zone, and 
T[\ion{S}{3}] in combination with the G92 relationship is used for the high ionization zone.  
In the cases with only one temperature measurement, 
the temperatures in the other zones are inferred from the G92 relationships.
Since we have [\ion{S}{3}] or [\ion{N}{2}] temperatures for 43 of the 45 \ion{H}{2} regions, 
we can adopt this relatively homogeneous method for nearly the entire dataset. 
We base the subsequent analysis on this prioritization.
For completeness, we present an abundance analysis based on T[\ion{O}{3}] in Appendix~\ref{sec:TO3}.

Ionic abundances relative to hydrogen are calculated using:
\begin{equation}
	{\frac{N(X^{i})}{N(H^{+})}\ } = {\frac{I_{\lambda(i)}}{I_{H\beta}}\ } {\frac{j_{H\beta}}{j_{\lambda(i)}}\ }.
	\label{eq:Nfrac}
\end{equation}
The emissivity coefficients, $j_{\lambda(i)}$, which are functions of both 
temperature and density, are determined using a 5-level atom approximation 
with the updated the atomic data reported in Table~\ref{tbl4}. 

Total oxygen abundances (O/H) are calculated from the simple 
sum of O$^{+}$/H$^{+}$ and O$^{++}$/H$^{+}$.
The other abundance determinations require ionization correction 
factors (ICF) to account for unobserved ionic species. 
For nitrogen, we employ the common assumption that N/O 
= N$^{+}$/O$^{+}$ \citep{peimbert67}.
This allows us to directly compare our results with other studies in the literature.
\cite{nava06} have investigated the validity of this assumption, and  
concluded that it is valid at a precision of about 10\%.

Collisionally excited emission lines of sulfur, neon, and 
argon are also observed in many of our spectra.
For Ne we use a fairly straightforward ICF: ICF(Ne) = 
(O$^{+}$ + O$^{++}$)/O$^{++}$ \citep{crockett06}.
S and Ar present more complicated situations as both S$^{++}$ and S$^{+++}$
lie in the O$^{++}$ zone and Ar$^{++}$ spans both the O$^{+}$ and the O$^{++}$ zones.
\citet{thuan95} provide analytic ICF approximations for both S and Ar using 
the model calculations of photoionized \ion{H}{2} regions by \citet{stasinska90}.
We employ these ICFs from \citet{thuan95} to correct for 
the unobserved S$^{+3}$, Ar$^{+2}$, and Ar$^{+4}$ states:
\footnotesize{
\begin{align}
        \mbox{ICF(S)} & =  \frac{\mbox{S}}{\mbox{S}^{+} + \mbox{S}^{++}} \nonumber \\
        			       & = \big[0.013 + x\{5.10 + x[-12.78 + x(14.77 - 6.11x)]\}\big]^{-1} \label{eqn:ICFS} \\
       \mbox{ICF(Ar)} & =  \frac{\mbox{Ar}}{\mbox{Ar}^{++} + \mbox{Ar}^{+++}} \nonumber \\
       			       & = \{0.99 + x[0.091 + x(-1.14 + 0.077x)]\}^{-1} \label{eqn:ICFAR1} \\
                    	       & =  \frac{\mbox{Ar}}{\mbox{Ar}^{++}} = [0.15 + x(2.39 - 2.64x)]^{-1}, \label{eqn:ICFAR2} 
\end{align}}
\normalsize \\
where $x =$ O$^{+}$/O and the second Ar equation is used 
when [\ion{Ar}{4}] $\lambda4740$ emission is not observed.
Ionic and total abundances are listed for 45 \ion{H}{2} regions in NGC~628 in Table~\ref{tbl5}.

\subsection{Alpha-Element Relative Abundances}
\label{sec:alpha}

\begin{figure}
\epsscale{1.2}
\plotone{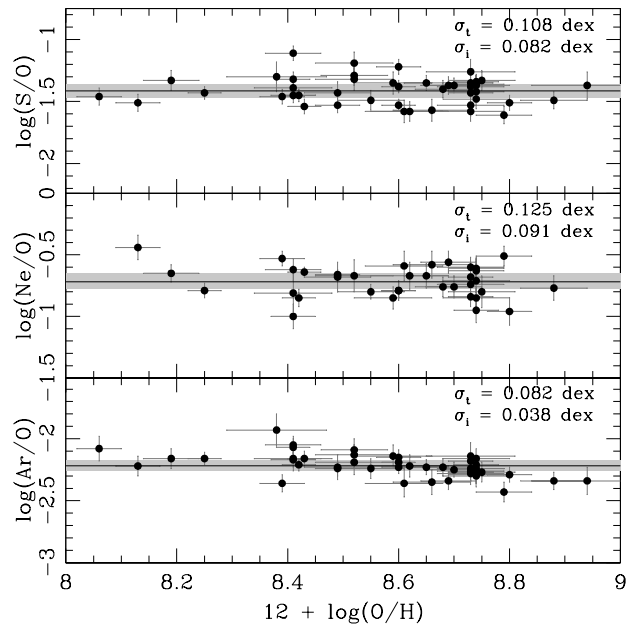}
\caption{$\alpha$-elements are plotted versus oxygen abundance for NGC~628.
The error-weighted average $\alpha$-element abundance is illustrated in each panel by a 
shaded box centered on the average and extended to $\pm1\sigma$.
The measured dispersions and calculated intrinsic dispersions are indicated 
in the upper right of each panel.
All three panels demonstrate constant $\alpha$-element/O ratio trends with small dispersions.}
\label{fig6}
\end{figure}

$\alpha$-elements are thought to be produced in massive stars such that 
sulfur, neon, and argon are produced in parallel with oxygen \citep[e.g.,][]{woosley95}.
Figure~\ref{fig6} shows the sulfur, neon, and argon abundances measured 
for NGC~628 versus oxygen abundance.
Error-weighted least-squares fits produce gradients within one standard 
deviation of a flat (zero) slope, and so we adopt the error-weighted average 
alpha-element abundances as constants, which are shown in Figure~\ref{fig6} 
as shaded boxes spanning one standard deviation.
The sample covers a range of 8.0 $<$ 12+log(O/H) $<$ 9.0 
with no obvious trends in sulfur, neon, or argon abundance.
S/O, Ne/O, and Ar/O all show very small intrinsic dispersions of 0.08, 0.08, and 0.02 dex, respectively.  
We report average $\alpha$-element to oxygen abundance ratios in Table~\ref{tbl6}.
In general, we find that the relative abundances of the $\alpha$-elements in NGC~628 behave as
expected for the nucleosynthetic products of massive stars.
The constant values over the range in metallicity are consistent with 
a universal value for the upper mass IMF.
This is in agreement with the recent study of the IMF in M31 clusters by \citet{weisz15}.

\subsection{Radial Gradients in N/O and O/H}
\label{sec:gradients}

\begin{figure}
\epsscale{1.2}
\plotone{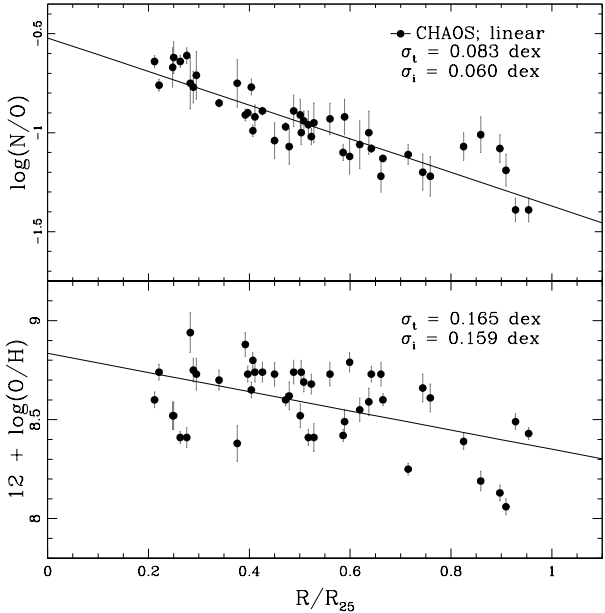}
\caption{N/O and O/H plotted versus galactocentric radius for NGC~628.
The solid lines represent the least-squares linear fits to the data when considering the errors in 
N/O and O/H abundance respectively. }
\label{fig7}
\end{figure}

We display the derived N/O and O/H abundances in 
Figure~\ref{fig7} as a function of galactocentric radius.
Because the locations of individual \ion{H}{2} regions are known
with high precision relative to one another, we consider only the uncertainties
associated with oxygen abundance here.
We characterize the N/O observations with a single, linear, error-weighted least-squares fit (solid line): \\
\footnotesize{
\begin{align} 
        & \log(\mbox{N/O}) \nonumber \\ 
        & = (-0.523\pm0.034) + (-0.077\pm0.006)\times{R}\  \mbox{ (dex/kpc)} \nonumber \\
        & = (-0.521\pm0.035) + (-0.849\pm0.064)\times{R}\  \mbox{ (dex/}R_{25}), \label{eqn:NO} 
\end{align} } 
\normalsize \\
with a measured dispersion in log(N/O) of $\sigma$ = 0.083 dex.  
Although this dispersion is small, the statistical analysis shows that most of this
dispersion is intrinsic ($\sigma_i$ = 0.060 dex).  

The bottom panel of Figure~\ref{fig7} shows the values of O/H as a function of galactocentric radius.
The resulting best fit to the 45 objects in the sample is given by: \\
\footnotesize{
\begin{align}
        & 12 + \log(\mbox{O/H}) \nonumber \\	
        & = (8.834\pm0.069) + (-0.044\pm0.011)\times{R}  \mbox{ (dex/kpc)} \nonumber \\
        & = (8.835\pm0.069) + (-0.485\pm0.122)\times{R} \mbox{ (dex/}R_{25}), \label{eqn:OH} 
\end{align} } 
\normalsize \\       
with a dispersion in log(O/H) of $\sigma$ = 0.165 dex.
Accounting for the observational uncertainties results in an intrinsic dispersion of 0.159 dex.
This is significantly larger than the dispersions seen in the relative abundances.
Since it is conceptually difficult to produce significant dispersions in absolute abundances
without significant dispersions in relative abundances (due to time delays of delivery and
mixing timescales), it is possible that this relatively large dispersion 
could be due to discrepant temperatures.
The final CHAOS gradients are tabulated in Table~\ref{tbl6}.


\begin{deluxetable*}{ccclcc}
\tabletypesize{\footnotesize}
\tablewidth{0pt}
\tablecaption{Adopted Abundance Gradients for NGC~628}
\tablehead{
\multicolumn{1}{c}{y} & \CH{} 	& \CH{x} 	& \CH{Equation of Correlation}  & \CH{$\sigma_t$}	& \CH{$\sigma_i$}	}
\startdata
{12+log(O/H)} 	& {vs.}	& {$R$ (dex/kpc)}		& {y = (8.834$\pm$0.069) + (-0.044$\pm 0.011)\times$X}	& {0.165} 	& {0.159}	\\
{}			& {}		& {$R$ (dex/$R_{25}$)}	& {y = (8.835$\pm$0.069) + (-0.485$\pm0.122)\times$X}  	& {}		& {}		\\
{log(N/O)}		& {vs.}	& {$R$ (dex/kpc)}		& {y = (-0.523$\pm$0.034) + (-0.077$\pm 0.005)\times$X}	& {0.083} 	& {0.060} 	\\
{}			& {}		& {$R$ (dex/$R_{25}$)}	& {y = (-0.521$\pm$0.035) + (-0.849$\pm0.064)\times$X}	& {}		& {}		\\
{log(S/O)} 		& {vs.} 	& {12+log(O/H)}		& {y = -1.413}									& {0.108} 	& {0.082} 	\\
{log(Ne/O)} 	& {vs.} 	&{12+log(O/H)}			& {y = -0.714}									& {0.125} 	& {0.091} 	\\
{log(Ar/O)} 	& {vs.} 	& {12+log(O/H)}		& {y = -2.219}									& {0.082} 	& {0.038} 	\\
\enddata
\tablecomments{The adopted best fits to the abundance gradients measured for the 3$\sigma$ NGC~628 CHAOS data.}
\label{tbl6}
\end{deluxetable*}


\section{DISCUSSION}


\subsection{Discrepant Temperature Measurements}\label{sec:discrepant}

\subsubsection{Suspect [\ion{O}{3}] Temperatures}\label{sec:badO3}

Historically, [\ion{O}{3}] $\lambda4363$ is the best studied direct temperature auroral line,
owing to its relative ease of observation, abundance of emitting ions, and 
increasing strength in the moderate- to low-metallicity \ion{H}{2} regions of nearby dwarf 
and (all but the central regions of) spiral galaxies.
Other common temperature-sensitive ratios, such as 
[\ion{N}{2}] I($\lambda\lambda6548,6584$)/I($\lambda5755$), 
[\ion{O}{2}] I($\lambda\lambda7320,7330$)/I($\lambda\lambda3726,3628$), and 
[\ion{S}{3}] I($\lambda\lambda9069,9532$)/I($\lambda6312$), 
are limited by their weaker abundances, origination in the smaller 
low-ionization zone, and combination of optical and near-infrared line ratios.
Because of these factors, the intensity ratio [\ion{O}{3}] $\lambda4363$/$\lambda\lambda$(4959+5007) 
has long been the chief diagnostic of nebular gas temperature.

In theory, the temperature measurements shown in Figure~\ref{fig5}
should fall along the lines of the relationships from the photoionization 
models of G92, but in many regards they do not.  
\citet{kennicutt03a} raised concerns about T[\ion{O}{2}], and additional 
studies have pointed out other significant temperature discrepancies 
\citep[e.g.,][]{bresolin05,pilyugin09,esteban09,zurita12,binette12}.
For the CHAOS data, Figure~\ref{fig5} demonstrates a surprisingly large dispersion for 
T[\ion{S}{3}] versus T[\ion{O}{3}], yet a tight relationship for T[\ion{N}{2}] versus T[\ion{S}{3}]. 
We investigate these two relationships further here.

\begin{figure*}
\begin{tabular}{cc}
	\includegraphics[scale = 0.425, trim = 0 0 0mm 0, clip]{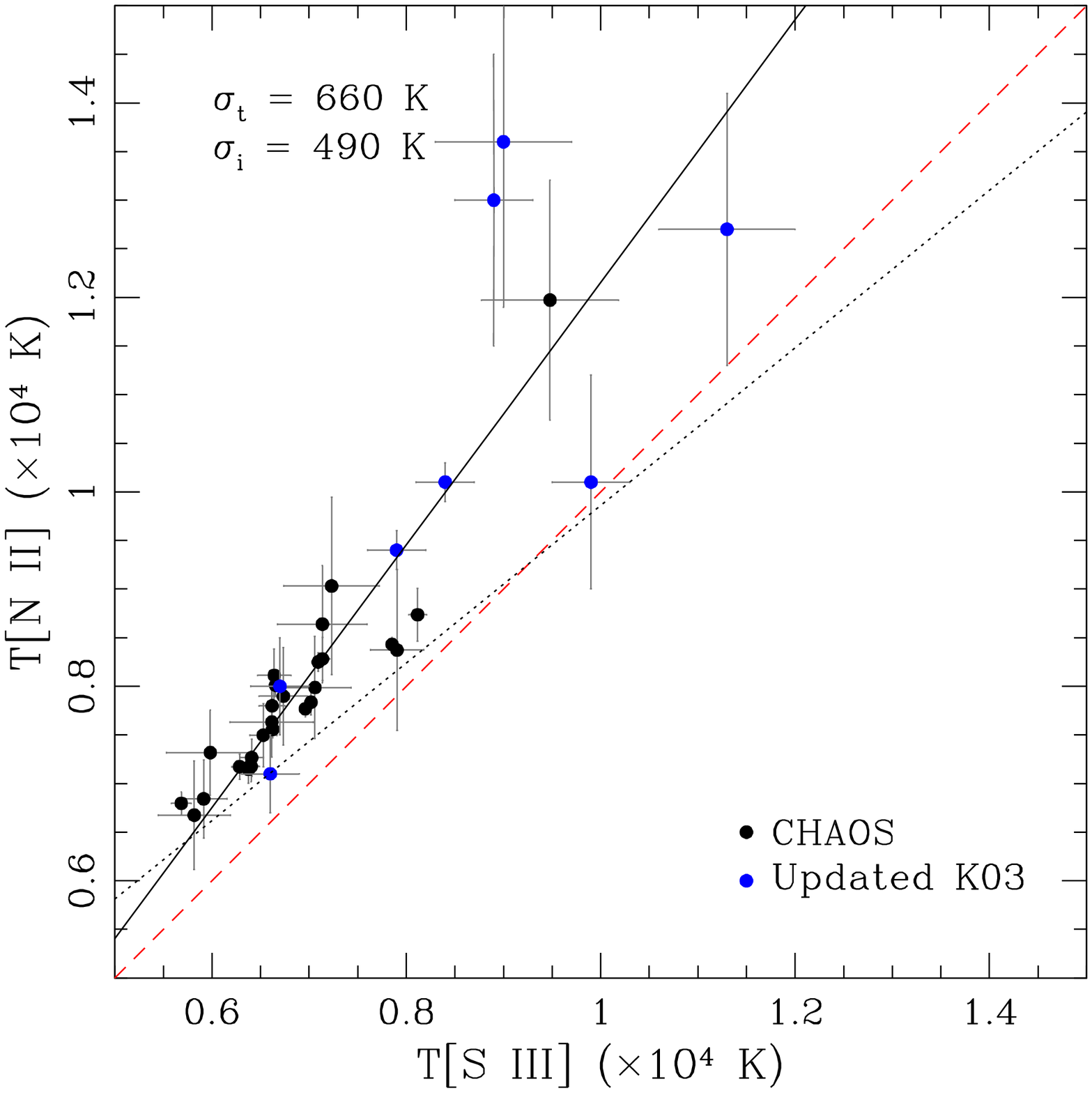} 
     & \includegraphics[scale = 0.425, trim = 0mm 0 0 0, clip]{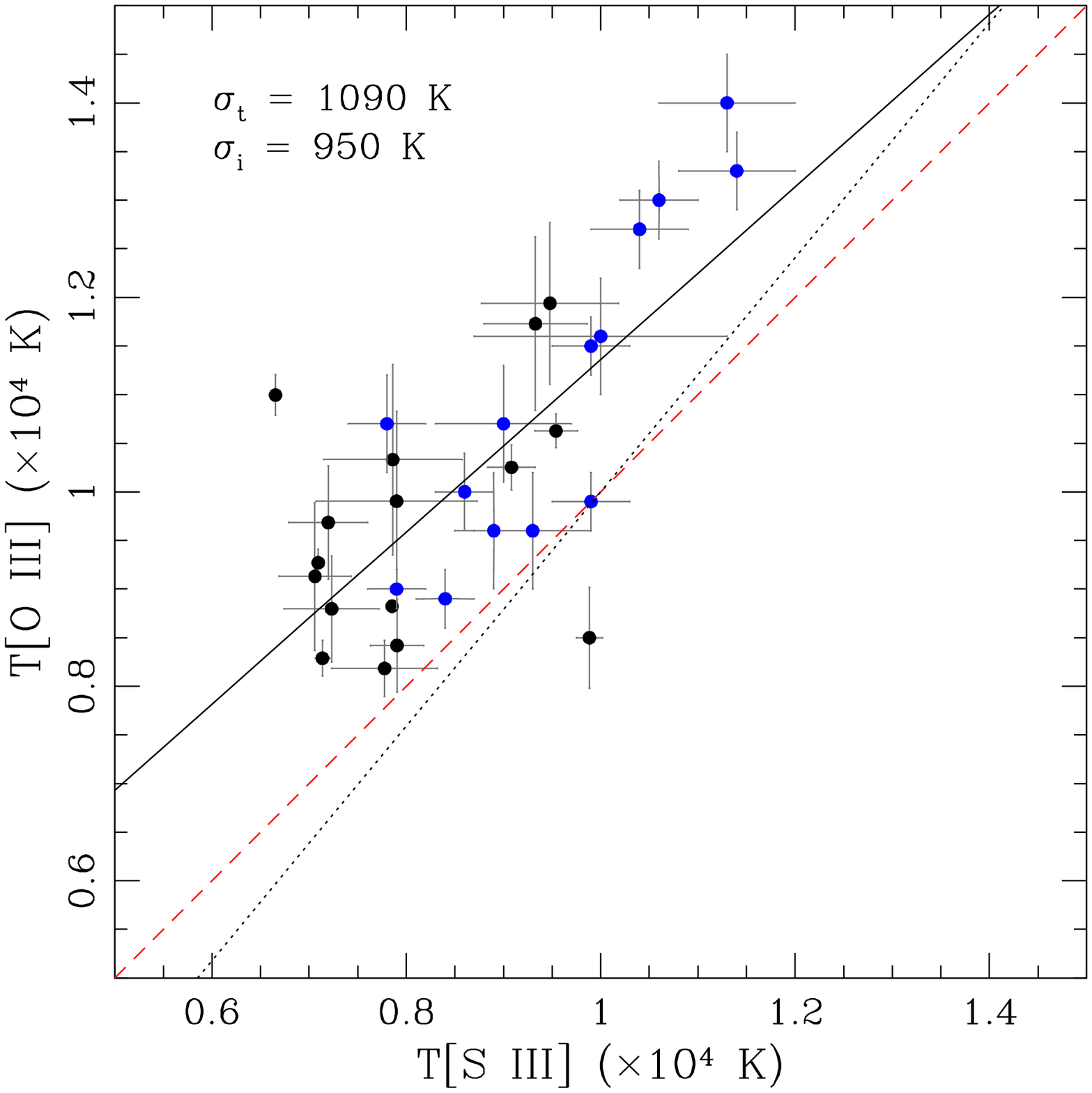}
\end{tabular}
\caption{Comparisons of the relations between T[\ion{N}{2}] and T[\ion{O}{3}] versus T[\ion{S}{3}].
The lines and solid points are the same as in Figure~\ref{fig5}.
A parallel comparison is provided by the M101 direct measurements of \citet{kennicutt03a},
where the open blue points correspond to the temperature values presented in their work 
and the closed blue points represent electron temperatures recalculated from their 
emission line fluxes with the adopted updated atomic data.}
\label{fig8}
\end{figure*}

We reproduce the CHAOS T[\ion{N}{2}]$-$T[\ion{S}{3}] data in the left panel of Figure~\ref{fig8}.
As mentioned earlier, the dispersion in this relationship about the line of equality is small.
K03 not did see such a tight relationship, but there were only 7 \ion{H}{2} regions with 
measurements of both T[\ion{N}{2}] and T[\ion{S}{3}].
The K03 observations have been added to Figure~\ref{fig8} for comparison\footnotemark[17].
For consistency, the K03 temperatures are recalculated using 
the K03 emission line fluxes and our adopted atomic data.
The least-squares fit to the combined set sees an increase in the dispersion 
due to the additional scatter at higher temperatures.
A very tight relationship between these two temperature diagnostics 
has been noted before \citep[e.g.,][]{bresolin05,zurita12, binette12}.
The emergence of an excellent correlation indicates that T[\ion{N}{2}] 
and T[\ion{S}{3}] provide highly consistent measurements of \ion{H}{2} 
region temperatures and increases confidence in their reliability at low temperatures.

In the right panel of Figure~\ref{fig8} we plot the CHAOS 
T[\ion{O}{3}]$-$T[\ion{S}{3}] measurements and compare them to the K03 data.  
K03 found excellent agreement between T[\ion{O}{3}]$-$T[\ion{S}{3}] measurements 
and the expectation from the photoionization modeling of G92.  
Our observations of NGC~628, coupled with updated atomic data,
are at odds with the M101 study, showing a significant dispersion
in the T[\ion{O}{3}]$-$T[\ion{S}{3}] comparison. 
The least-squares fit to the combined data set results in a total dispersion of 
1090\,K which corresponds to an intrinsic uncertainty of 950\,K.
As shown in Figure~\ref{fig3}, the original K03 [\ion{S}{3}] temperatures are higher by about 10\%.
The result is a shift in the T[\ion{O}{3}]$-$T[\ion{S}{3}] relationship away from 
the G92 relationship, and is equivalent to the offset observed for the CHAOS data.
\citet{perez-montero06} suggest a theoretical T[\ion{O}{3}]$-$T[\ion{S}{3}] 
relationship that also differs from the G92 standard, identifying 
new atomic coefficients as the source of the difference.
However, the new atomic data have minimal effect on the observed dispersions.
One very important thing to note from Figure~\ref{fig8} is the difference
in the temperature ranges covered by the observations in the two plots.
The left hand panel concentrates on a tight correlation for values of T[\ion{N}{2}] below 
$\sim$ 9,000 K, while the bulk of the T[\ion{O}{3}] values in the right hand panel, which show
significant scatter, are above $\sim$ 7,000 K.
Thus, the conclusions from K03 and the present study could be understood as an increasing 
dispersion in the T[\ion{O}{3}]$-$T[\ion{S}{3}] relationship with decreasing temperature.  
This result appears to be supported by the right hand panel of Figure~\ref{fig8}.

\footnotetext[17]{In order to assure the quality of the data compared and uniform 
methodology, a 3$\sigma$ S/N cut in flux was made for the M101 data.}


\subsubsection{Sources of Temperature Discrepancies}

We now turn to the possible causes of the significant temperature discrepancies
demonstrated by Figures~\ref{fig5} and \ref{fig8}.
Due to the moderately-high resolution of MODS, 
[\ion{O}{3}] $\lambda4363$ is resolved from the Hg $\lambda4358$ sky line.
Additionally, we examine the [\ion{O}{3}] auroral line emission by hand in both
the one-dimensional and two-dimensional spectra, eliminating concerns of
cosmic rays, Balmer absorption, or other measurement errors. 
Recently, \citet{binette12} have highlighted the discrepancy between
the T[\ion{O}{3}] and T[\ion{S}{3}] temperatures
and investigated possible sources of this disagreement:
temperature inhomogeneities, metallicity inhomogeneities, a non-Maxwell-Boltzmann 
distribution of electron energies, and shocks.
Temperature inhomogeneities within individual \ion{H}{2} regions were first proposed 
by \citet{peimbert67}, where non-uniform temperature distributions tend to result 
in average observed temperatures biased higher than true local temperatures.  
However, \citet{binette12} found that simple temperature fluctuations are insufficient 
to account for the magnitude of the T[\ion{O}{3}]$-$T[\ion{S}{3}] disparity, especially 
at the low electron temperatures characteristic of high abundance \ion{H}{2} regions 
(see their Figure 4).
Instead, Binette et al.\ conclude that metallicity inhomogeneities 
\citep[e.g.,][]{tsamis03,tsamis11}, adopting a $\kappa$-distribution 
for the electron energy distribution \citep{nicholls12}, and pollution 
by shock waves all successfully reproduce the observed T[\ion{O}{3}] 
excesses (see their Figures~$5-8$).
However, \citet{mendoza14} found $\kappa$-distributions only account for observed temperature 
variations in low-excitation regions, and do not reconcile the differences in high-excitation regions.

For the low electron temperatures associated with the CHAOS data,
\citet{binette12} demonstrate that plane-parallel shock models which propagate within a 
photoionized gas layer successfully bridge the T[\ion{O}{3}]$-$T[\ion{S}{3}] gap.
The shocked nebula models show a diminishing excess of T[\ion{O}{3}] with 
decreasing nebular metallicity (increasing T) similar to what we see in our data, 
where the largest excesses occur with decreasing temperature (increasing metallicity)
towards the inner disk.
By combining the continuous starburst model of \citet{kewley01} and the 
shock+precursor model of \citet{allen08}, \cite{binette12} suggest an 
additional region which is a mixing zone of the photo- and shock-ionization zones.

\begin{figure}
\epsscale{1.2}
\plotone{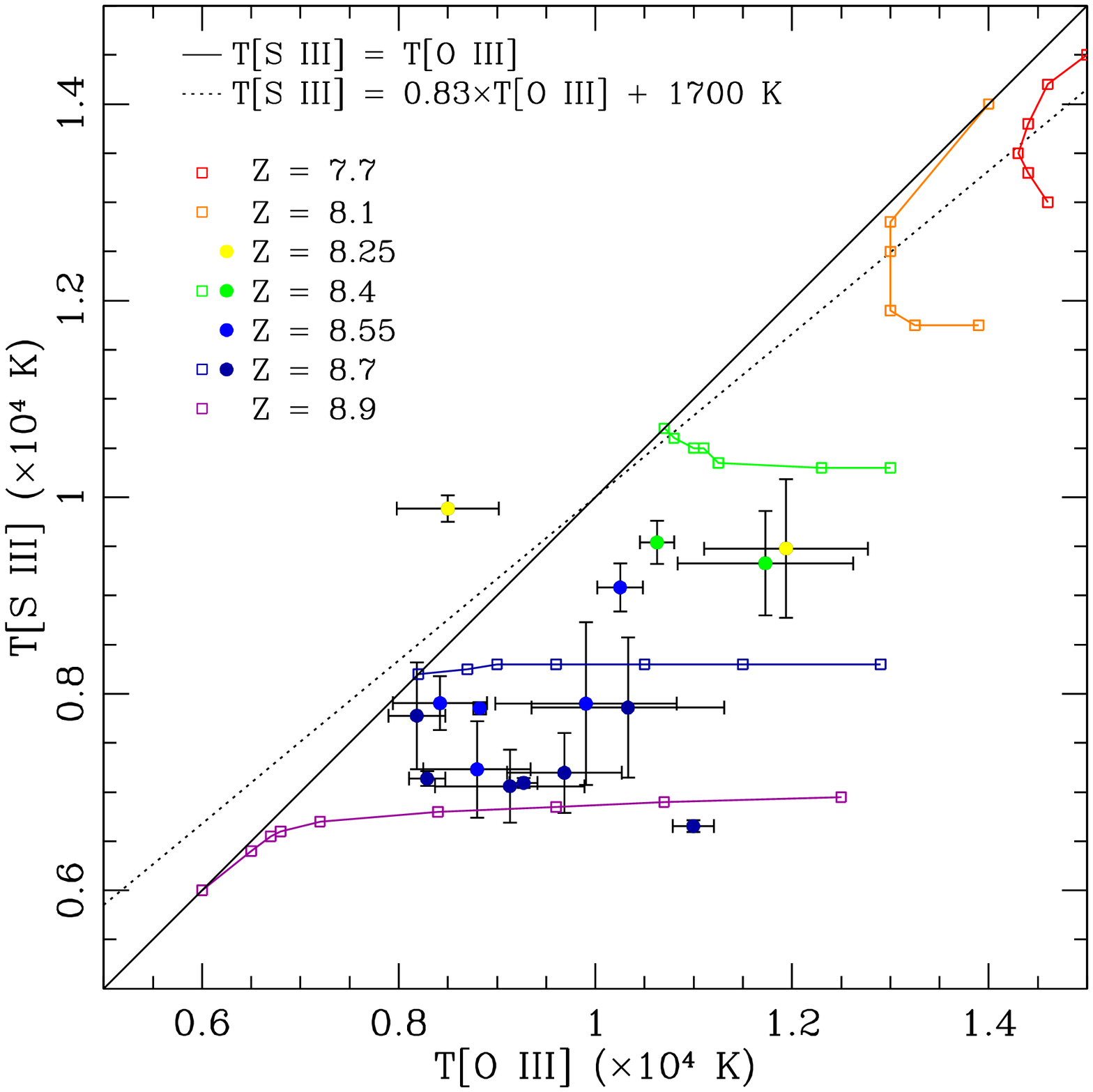}
\caption{T[\ion{O}{3}] versus T[\ion{S}{3}] plot for NGC~628 in comparison
to models including the effects of shocks from \citet{binette12}.
Abundances are divided into bins and points are colored coded by
the central oxygen abundance of each bin.
Only points with measurements of both [\ion{S}{3}] and [\ion{O}{3}] temperatures
are plotted as solid circles.
The models of \citet{binette12}, which combine the continuous starburst model of \citet{kewley01} 
and the shock+precursor model of \citet{allen08}, are plotted as open squares.}
\label{fig9}
\end{figure}

In Figure~\ref{fig9} we reproduce the shock model data from Figure 8 of 
\citet{binette12} and compare the CHAOS data for NGC~628.
The solid points are the \ion{H}{2} regions which have both measured [\ion{O}{3}] and 
[\ion{S}{3}] temperatures, and are color coded by abundance.  
Points in this plot show that (with the exception of one major outlier: 
NGC~628-168.2+150.8) T[\ion{O}{3}] is in excess of T[\ion{S}{3}], where 
increased temperatures subsequently lead to artificially low oxygen abundances.
This idea is congruent with the picture shown in Figure~\ref{fig9}:
the colors of the points indicate that their abundances are higher than those 
of the models to which they are closest.
If their true photoionized T[\ion{O}{3}] is lower ($\sim$T[\ion{S}{3}] ), 
then the true oxygen abundance is
higher and in better agreement with the model abundances.

It is clear from studies in the literature and from our CHAOS observations
of NGC~628 that emission line spectra for some \ion{H}{2} regions are
affected by one or more processes which lead to discrepant temperatures
derived from auroral line observations.  
By comparing the present observations with theoretical models, we have 
shown that shocks, as an example, can contribute to discrepant temperatures; 
however, the true nature of the perturbations is not clear.  
It is interesting to note that the auroral lines with the highest excitation energies 
(e.g., [\ion{O}{3}] $\lambda4363$ and [\ion{O}{2}] $\lambda\lambda7320,7330$, 
with excitation energies of 5.3 and 5.0 eV respectively) show the largest 
discrepancies, while those with lower excitation energies (e.g., [\ion{N}{2}] 
$\lambda5755$ and [\ion{S}{3}]$\lambda6312$, with excitation  energies of 
4.1 and 3.3 eV respectively) are less sensitive to these perturbations.  
The higher excitation energies of the oxygen temperature diagnostic 
lines may contribute to making these lines more vulnerable to the 
perturbations, regardless of the physical cause.
This leads quite naturally to the result that the biases 
are larger in higher abundance (lower temperature) 
\ion{H}{2} regions, in agreement with the observations.
From the present NGC~628 dataset, we cannot deduce 
a single likely cause for the temperature discrepancies. 
However, in Appendix~\ref{sec:indicators} we suggest relative Ar abundances 
as excellent diagnostics for identifying which spectra are most affected.  
We will revisit this problem with the entire CHAOS dataset.


\subsection{Comparison of Gradients with Measurements from the Literature}
\label{sec:comps}

Although NGC628 is one of the best-studied spiral galaxies, 
there are only a handful of previous studies with direct abundance measurements.
We first perform a straightforward comparison to these studies, and then extend
this comparison to the integral field spectroscopy study by \citet{rosales-ortega11}.

Three previous studies present direct abundance measurements for NGC~628: 
\citet[][14 regions, hereafter B13]{berg13}, \citet[][7 regions, hereafter C13]{croxall13}, and 
\citet[][1 region]{castellanos02}, but only the first reports nitrogen or $\alpha$-element abundance ratios.
Note that the three studies derive electron temperatures using different methods.
B13 used the [\ion{O}{3}] and [\ion{N}{2}] auroral lines to determine electron temperatures, 
\citet{croxall13} used the infrared [\ion{O}{3}] $88$\micron\ fine-structure line, 
and \citet{castellanos02} found agreement amongst four temperature determinations 
from the [\ion{O}{3}], [\ion{S}{3}], [\ion{N}{2}], and [\ion{S}{2}] auroral lines observed in one \ion{H}{2} region.  
We have not attempted to make any corrections to place these
different abundance derivations on a common scale.

\subsubsection{Radial Gradients in N/O and O/H}
\label{sec:gradientscomp}

The measure of nitrogen across NGC~628 offers further 
insight into the nature of abundance gradients.
Because nitrogen and oxygen are formed primarily in intermediate- 
and high-mass stars respectively, they are returned to the interstellar 
medium on different time scales
\citep[e.g.,][]{garnett90,vanzee98a,vanzee98b,berg12}.
In the upper panel of Figure~\ref{fig10} we plot the CHAOS N/O gradient, 
including 11 observations with T[\ion{N}{2}] measurements 
and three observations with T[\ion{O}{3}] measurements from B13.
The values from the literature are consistent with the trends established by the CHAOS observations;
the main difference is the extension to greater radius afforded by the B13 points.
\citet{berg13} used a likelihood ratio F-test to show that the N/O 
gradient is best described by a bi-modal relationship (short dashed line), where 
decreasing N/O flattens out near the isophotal radius (R$_{25}\sim11$ kpc,).

In the lower panel of Figure~\ref{fig10} we illustrate the oxygen 
abundance relationship across the disk of NGC~628.
In comparison to the CHAOS data (solid points), direct abundances from B13 and C13 
are plotted as open squares and open triangles respectively. 
The reduced level of scatter amongst sources when prioritizing T[\ion{S}{3}] and T[\ion{N}{2}] 
is confirmed when incorporating the additional points from the literature.
Since the infrared fine-structure lines used by C13 are thought to produce 
abundances that are largely independent of electron temperature, 
the fact that these values lie in unison with the CHAOS data 
offer another confirmation of the chosen temperature determinations.
Further, B13 measured T[\ion{N}{2}] in 11 \ion{H}{2} regions to obtain 
directly comparable abundances, and these values agree with the 
dispersion in the CHAOS sample.
In short, this review of the oxygen abundance gradient with respect to a variety 
of temperature determinations supports our decision to conduct a homogeneous 
abundance analysis based on T[\ion{S}{3}] and T[\ion{N}{2}]. 


\subsubsection{Comparison with Results from Rosales-Ortega (2011)}
\label{sec:RO11}

The most comprehensive emission line abundance survey of NGC~628 
to date was carried out by \citet[][hereafter RO11]{rosales-ortega11}.
The RO11 study performed a nearly complete 2D sampling of the disk within 
the isophotal radius, allowing a variety of statistically significant analyses.
Complimentary to the work presented here, RO11 performed an in-depth comparison 
of various strong-line metallicity indicators and laid a framework of geometric parameters.

RO11 obtained robust 2D fiber spectroscopic observations of NGC~628, 
which were converted to a set of 96 \ion{H}{2} region spectra which are
directly comparable to the CHAOS observations.
With this data set, different strong-line abundance estimators were 
used to explore the effects of choosing different calibrations.
The RO11 ff-$T_e$ \citep{pilyugin05b} analysis of the \ion{H}{2} region catalog mimics the data collection 
and analysis presented here, and allows the following side-by-side comparison.

\begin{figure*}
\centering
{\textbf{CHAOS Abundance Gradients for NGC~628}\par\medskip}
\vspace*{-0.5cm}
\plotone{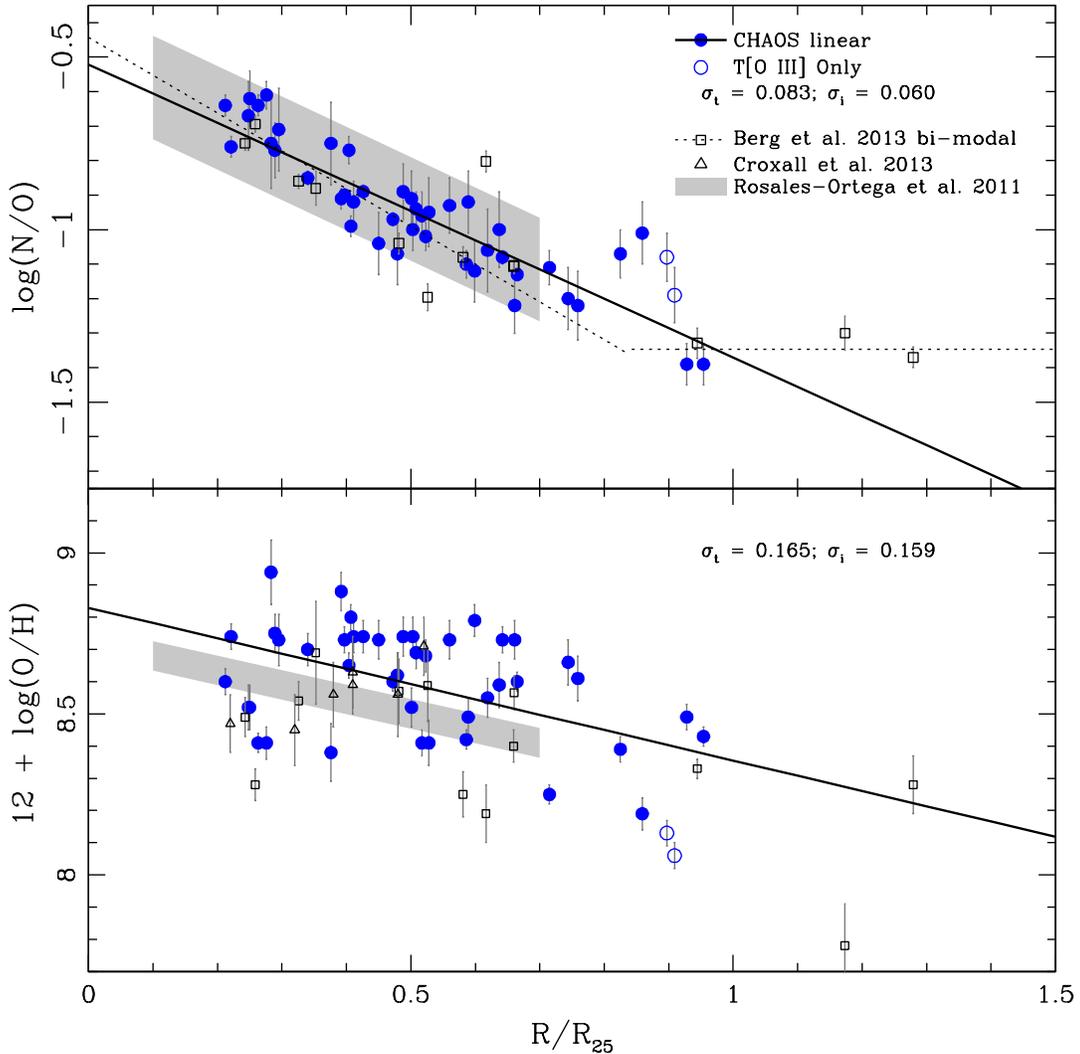}
\caption{N/O and O/H plotted versus galactocentric radius for NGC~628 and 
are compared to direct and ff-$T_e$ methods in the literature.
Data from B13 (open squares) and from C13 (open triangles)
are plotted, where electron temperatures were determined using [\ion{O}{3}] and/or
[\ion{N}{2}] for the former and the infrared fine-structure [\ion{O}{3}] $88\mu$ line for the latter.
The solid line depicts the least-squares best fits to the CHAOS data, 
as reported in Section~\ref{sec:abund}, whereas the shaded box 
represents the negative gradient and dispersion measured for the optical disk by RO11 using the ff-$T_e$ method.
In the top panel, the CHAOS data intersects the bi-modal N/O relationship observed by B13 (dashed line).
In the bottom panel, the [\ion{N}{2}] based B13 data and the nearly 
temperature independent measurements of C13 coincide with the CHAOS data.}
\label{fig10}
\end{figure*}

In Figure~\ref{fig10} we compare N/O and O/H radial gradients 
from the  ff-$T_e$ calibration of RO11.
The top panel of Figure~\ref{fig10} illustrates the general agreement between 
the direct abundance and ff-$T_e$ N/O gradients for the optical disk.
Within the uncertainty of measurement, the CHAOS best fit (Equation~\ref{eqn:NO}:
 log(N/O) = ($-0.523\pm0.034$) + ($-0.077\pm0.005)\times{R}$ (dex/kpc)) parallels 
the RO11 fit (log(N/O) = ($-0.50\pm0.03$) + ($-0.080\pm0.06)\times{R}$ dex/kpc)\footnotemark[18],
in particular, note that the gradients are within 1$\sigma$ of one another.
Since the N/O ratio has little dependence on temperature, 
the two methods provide nearly identical results except for a small average shift.
Note the outlined, shaded box displays the 0.3 dex 
spread in the measured data as reported by RO11.
Similarly, the bottom panel of Figure~\ref{fig10} 
compares these two methods for the O/H gradient.
With the addition of data from the literature extending into the outer disk (beyond 
$R_{25}$), RO11 propose a tri-modal oxygen abundance gradient: a 
steeply declining abundance gradient for the majority of the optical disk, 
with flatter features within the inner 2 kpc and in the outer disk.
The mid O/H slope ($-0.034\pm0.02$ dex/kpc) from RO11, extending from $\sim0.1-0.7$ 
R/R$_{25}$, is shown as a grey shaded box in Figure~\ref{fig10} whose width 
represents the 1$\sigma$ error-weighted dispersion in the log(O/H) measurements.  
For this part of the optical disk, the RO11 results are in agreement with the
slope from the CHAOS observations ($-0.044\pm0.011$ dex/kpc),
but are offset lower by $\sim0.1$ dex.
The CHAOS observations do not cover the inner or 
outer regions where RO11 find different slopes.

\footnotetext[18]{RO11 use a distance of 9.3 Mpc to deproject radial distances,
in contrast to the 7.2 Mpc assumed here. Since the most reliable distance determinations,
such as the TRGB distance, have not been performed for NGC~628, this choice is simply one
of preference. Here we use the gradient fits in terms of luminosity distance from RO11
and convert them to match the scale of the CHAOS relationships.}


\subsection{Dispersions in Abundances at a given Galactocentric Radius}

The dispersion in abundances within the ISM of a galaxy is a very important observable, but 
there are very few galaxies with a sufficient number of observations to make statistically
significant conclusions.  Based on 61 direct [\ion{O}{3}] $\lambda4363$ measurements,
\citet[][hereafter RS08]{rosolowsky08} determined a significant dispersion in the oxygen 
abundance gradient of M33 ($\sigma_i = 0.11$ dex).
In this study, a number of parameters, including L(H$\alpha$), c(H$\beta$), Balmer absorption
and emission equivalent widths, and ionization correction factor, were tested for correlated against the derived
metallicity gradient and its residuals, but no trend indicative of systematic errors was found.
The findings of RS08 were challenged by \citet{bresolin11} using 8 observations of 
the [\ion{O}{3}] $\lambda4363$ auroral line and the $R_{23}$ strong-line method.
Based on this analysis, \citet{bresolin11} measured a much smaller dispersion, and speculated 
that the significant scatter seen in the $T_e$-based abundances of M33 from the RS08 
sample is due to measurement errors in the weak [\ion{O}{3}] $\lambda4363$ auroral line.

Although relative abundances from the CHAOS observations show very 
small dispersions at a given galactocentric radius in NGC~628, the 
absolute oxygen abundances show a large and significant dispersion.
RO11 also measured a large dispersion in metallicity, finding the scatter to be 
larger in their fiber-to-fiber sample (rms = 0.128 dex) than the \ion{H}{2} region 
catalog (rms = 0.070 dex) (where several fiber spectra are averaged together). 
This level of spread is consistent with the CHAOS sample ($\sigma_t = 0.165$ dex, $\sigma_i = 0.159$ dex).

\begin{figure}
\begin{tabular}{c}
	\includegraphics[scale = 0.45, trim = 0mm 0mm 10mm 100mm, clip]{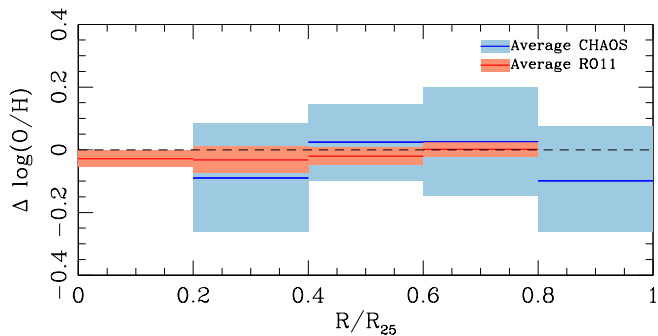}
\end{tabular}
\caption{Average offset in oxygen abundance from the measured radial gradient plotted 
in radial bins of 0.2 dex for each the CHAOS sample and the RO11 study.
Colored solid lines represent the average offset in each bin, where the extent of the shaded box
depicts the dispersion around the mean offset.}
\label{fig11}
\end{figure}

In Figure~\ref{fig11}, we plot the average offset in O/H abundance from the best fit relationship
in radial bins of 0.2 $R/R_{25}$ for both the CHAOS sample and the RO11 study.
For the three bins in common between $0.2-0.8$ $R/R_{25}$, the RO11 data has both smaller
average departures from the measured radial gradient and smaller dispersions around these mean offsets.
This result further suggests that the large dispersion measured in the CHAOS gradient 
has a physical origin whose effect is removed when averaged over entire \ion{H}{2} regions.
Be that as it may, until the source(s) of the discrepant temperature measurements are better understood, 
at this point, it is best to consider the measured dispersion to be an upper limit.

In addition to temperature discrepancies, we investigate other sources of scatter in oxygen abundance, 
such as the symmetry of the metal abundance distributions, but find no obvious relationship with spatial structure 
(differences amongst spiral arms or arm versus inter-arm abundances) or ionization fraction.
One advantage of the RO11 study over CHAOS is that it samples a nearly complete coverage of the disk.
Since we have shown that the two studies find comparable oxygen abundance gradients, 
we can use the RO11 disk geometry analysis to complement our understanding of the CHAOS data.
RO11 performed two detailed spatial analyses: (1) by quadrants and (2) by spiral arms, as defined by 
\citet{kennicutt80}.
RO11 found distinct radial trends in the ionization parameter and 
metallicity for both the geometrical quadrants and the morphological arms, 
indicating that the disk may have evolved somewhat asymmetrically.
However, these variations result in O/H gradients which lie within the 
uncertainties of the overall abundance trend for the disk.

While the CHAOS and RO11 studies suggest the significant intrinsic oxygen abundance dispersion 
found for NGC~628 is real, a definitive source for such variation at a given radius has not been found.
Further, considering the the contrasting results of the RS08 and \citet{bresolin11} studies for M33, 
we speculate that these azimuthal variations are simply not recovered in small samples.
For the large samples of direct abundances presented here and in RS08, 
the significant intrinsic dispersions measured are due to the electron temperatures.
However, the dispersion could be due to discrepant temperatures, uncertainty in the measurements, or both. 
Following the line of argument laid out in \citet{binette12} and summarized in Section~\ref{sec:discrepant},
the primary factor influencing temperature discrepancies, and subsequently determined abundances, is not yet clear.


\section{SUMMARY AND CONCLUSIONS} \label{sec:sum}

The CHemical Abundances of Spirals (CHAOS) project seeks to establish a broader 
understanding of the chemical evolution of spiral galaxies in general.
CHAOS harnesses the combined power of the Large Binocular Telescope (LBT) with the large 
spectral range and sensitivity of the Multi Object Double Spectrographs (MODS) to
observe ``direct" abundances in a large sample of \ion{H}{2} regions in spiral galaxies.
In this manner, we measure the largest sample of the highest quality spectra to date in NGC~628,
with one or more temperature-sensitive auroral lines
([\ion{O}{3}] $\lambda4363$, [\ion{N}{2}] $\lambda5755$, and [\ion{S}{3}]$\lambda6312$) 
being observed at a strength of 3$\sigma$ or greater in 45 \ion{H}{2} regions.
Analysis of this data set provides the following insights:

1) Due to the obvious advantages in observing [\ion{O}{3}] $\lambda4363$ over the 
other auroral lines, it has long been the principle nebular temperature diagnostic.
Further, the scarcity of observations of other lines in the literature 
has postponed any comparative tests of validity.
The CHAOS sample presented here records multiple auroral lines in many of the \ion{H}{2} 
regions observed, allowing a comparison of derived temperatures from multiple emission lines.
We have shown a very tight relationship between temperatures based on [\ion{S}{3}] $\lambda6312$
and [\ion{N}{2}] $\lambda5755$ indicating that these temperatures are reliable. 
In contrast, temperatures from [\ion{O}{2}] $\lambda\lambda7320,7330$ and 
[\ion{O}{3}] $\lambda4363$ can often show large discrepancies compared to T[\ion{S}{3}] and T[\ion{N}{2}].

2) These results lead us to adopt a uniform method of electron temperature determination
in which T[\ion{S}{3}] and T[\ion{N}{2}] are prioritized and used in combination
with the theoretical relationships of G92.
Through this approach, we were able to minimize the dispersion seen in the abundance gradients.  

3) Relative abundances of S/O, Ne/O, and Ar/O, which are nearly independent of 
temperature, are measured to be constant across the entire optical disk of the galaxy.
Using weighted least-squares fits, with errors in both the x and y variables, 
all three show very small intrinsic dispersions (0.08, 0.09, and 0.04 respectively).
These unvarying values over the range in metallicity are consistent with a universal
value for the upper mass range of the IMF.

4) Radial gradients are measured for both N/O and O/H out to the optical radius of NGC~628.
The N/O ratios show a negative gradient, consistent with the previous works of 
\citet{berg13} and \citet{rosales-ortega11}, but do not sample the radial extent
necessary to assess the presence of a bi-modal relationship.
The small measured dispersion about the N/O relationship is attributed to 
intrinsic dispersion ($\sigma_t = 0.083$; $\sigma_i = 0.060$).
We interpret this result as an indication that the ISM in NGC~628 is
chemically well mixed. 

5) O/H abundances show a moderate negative radial gradient of $-0.485\pm0.122$ 
dex/R$_{25}$ and a significant dispersion of $\sigma_t = 0.165$ dex, in line with the 
previous work of \citet{rosales-ortega11}.
We posit that this dispersion represents an upper limit to the true
dispersion in O/H at a given radius.
The dispersion in the N/O gradient is significantly smaller than in the O/H gradient.
Since N/O is nearly temperature independent, some of the scatter seen in the 
O/H relationship is likely 
due to systematic uncertainties arising from the temperature measurements.

6) Currently, typical analyses of \ion{H}{2} region abundances 
assume a simple symmetric, 3-zone model.
The electron temperatures determined for these zones show large discrepancies,
but their physical nature has not been resolved. 
Factors such as shocks, temperature inhomogeneities, abundance 
inhomogeneities, and non-Maxwellian electron energy distributions 
may each play a role. 
Improvements in atomic data, photo- and shock-ionization models, and further 
high-quality, statistically significant observations are needed to address this issue.


\acknowledgements
DAB and KVC are grateful for support from NSF Grant AST- 6009233.
DAB is also grateful for support from the Dissertation Fellowship from the University of Minnesota.
We thank the anonymous referee who provided very useful commentary on the atomic data.

This paper uses data taken with the MODS spectrographs built with funding from 
NSF grant AST- 9987045 and the NSF Telescope System Instrumentation Program (TSIP), 
with additional funds from the Ohio Board of Regents and the Ohio State University Office of Research. 
This paper made use of the modsIDL spectral data reduction pipeline developed by KVC in part with funds provided by NSF Grant AST-1108693.
This work was based in part on observations made with the Large Binocular Telescope (LBT). 
The LBT is an international collaboration among institutions in the United States, Italy and Germany. 
The LBT Corporation partners are: the University of Arizona on behalf of the Arizona university system; 
the Istituto Nazionale di Astrofisica, Italy; the LBT Beteiligungsgesellschaft, Germany, representing the 
Max Planck Society, the Astrophysical Institute Potsdam, and Heidelberg University; the Ohio State University; 
and the Research Corporation, on behalf of the University of Notre Dame, the University of Minnesota, 
and the University of Virginia. 

This material is based upon work supported by the National Science Foundation under Grant No. 1109066
and has made use of NASA's Astrophysics Data System
Bibliographic Services and the NASA/IPAC Extragalactic Database
(NED), which is operated by the Jet Propulsion Laboratory, California
Institute of Technology, under contract with the National Aeronautics
and Space Administration.


\appendix


\section{OXYGEN ABUNDANCES USING T[\ion{O}{3}]}
\label{sec:TO3}

We discussed the discrepancies between [\ion{S}{3}] and [\ion{O}{3}] electron temperatures
in Section~\ref{sec:badO3}, and charted them in Figure~\ref{fig8}. 
The uniform analysis approach adopted in this paper prioritizes calculations of the
high ionization zone temperature from T[\ion{S}{3}] over T[\ion{O}{3}].
Here we offer an additional validation of this choice.

\begin{figure}
\plotone{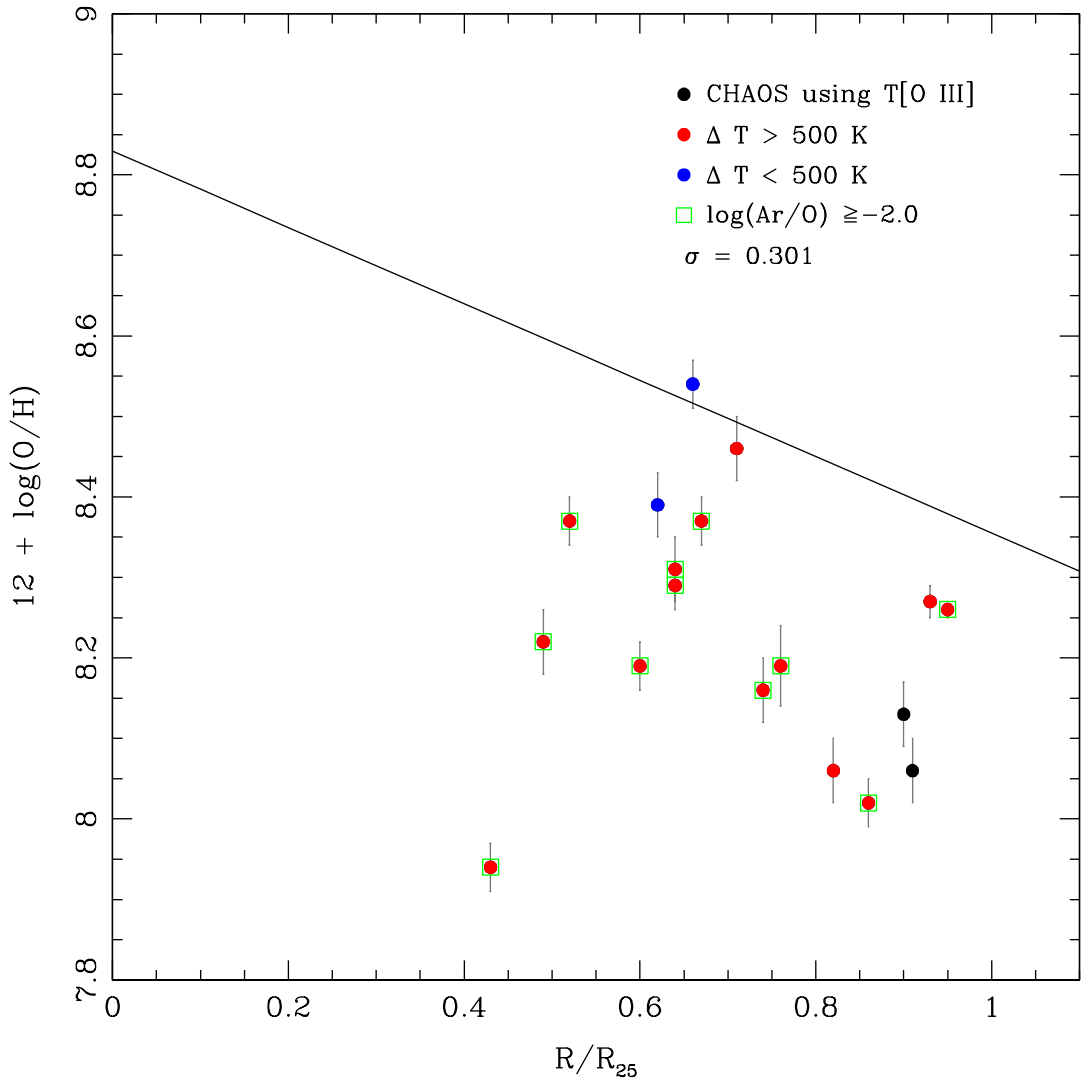}
\caption{The oxygen abundance gradient for NGC~628 using only [\ion{O}{3}] for temperature determinations
and the subsequent abundance analysis.
The best least-squares fit determined in Section~\ref{sec:gradients} is compared as a solid line,
and the dispersion about this relationship is indicated.
Regions used in the temperature discrepancy analysis are denoted as colored points,
where the highly discrepant points are red and non-discrepant points are blue.
Log(Ar/O) outliers are outlined by green boxes.}
\label{fig12}
\end{figure}

In Figure~\ref{fig12}, for the 18 regions with [\ion{O}{3}] auroral line detections, 
we plot the oxygen abundance gradient adopting [\ion{O}{3}] as the standard electron temperature indicator. 
The solid black line is the error-weighted least-squares gradient (Equation~\ref{eqn:OH}) 
that we determined in Section~\ref{sec:gradients}.
Regions with both [\ion{O}{3}] and [\ion{S}{3}] temperature determinations are color coded
according to magnitude of the temperature discrepancy, and log(Ar/O) outliers 
(as discussed in Appendix~\ref{sec:indicators}) are outlined by green boxes.
Several trends are notable in this plot:
(1) the scatter in oxygen abundance seems to be systematically shifted to lower abundances,
(2) the greatest deviants from the gradient have large temperature discrepancies, and
(3) the dispersion relative to the best fit is large ($\sigma_t$ = 0.301 dex).
Thus, the adoption of our analysis approach (see Section~\ref{sec:method}) 
resulted in a a significant reduction in the scatter ($\sigma_t$ = 0.165 dex).
Because Ar and S abundances are sensitive to an intermediate temperature, 
a propagation of discrepancies ensues if [\ion{S}{3}] is in disagreement with T[\ion{O}{3}] and T[\ion{N}{2}]. 
Minimizing these temperature inconsistencies removed many of the Ar/O, S/O, and O/H disparities seen in our data.

\section{INDICATORS OF TEMPERATURE DISCREPANCIES}
\label{sec:indicators}

[\ion{O}{3}] has been and continues to be a useful temperature indicator, 
but significant deviations have emerged such that multiple cases must be considered, 
and the appropriate modifications applied:
(1) In the case that T[\ion{O}{3}]$\sim$T[\ion{S}{3}], either indicator or an average may be adopted.
(2) When T[\ion{S}{3}]-T[\ion{O}{3}] discrepancies are large, the multiple-line 
diagnostics presented throughout this paper indicate T[\ion{S}{3}] should be used. 
(3) When T[\ion{O}{3}] is the only measured temperature some risk of overestimating the temperature 
may be introduced, but in a systematic way such that oxygen abundance is underestimated.

In the last case of only one temperature determination, there is no potentially better alternative.
However, if multiple auroral line detections are present, a further indication of how to proceed is needed.
In this section we present two such indicators that appear promising as possible temperature diagnostics.
We consider the standard choice of temperature indicators: T[\ion{O}{3}] for the high ionization zone,
T[\ion{S}{3}] for the intermediate ionization zone, and T[\ion{N}{2}] for the low ionization zone.
The following plots are intended as an introduction for analysis to come with the completed CHAOS dataset,
and so no conclusions are drawn here.

\begin{figure}
\plotone{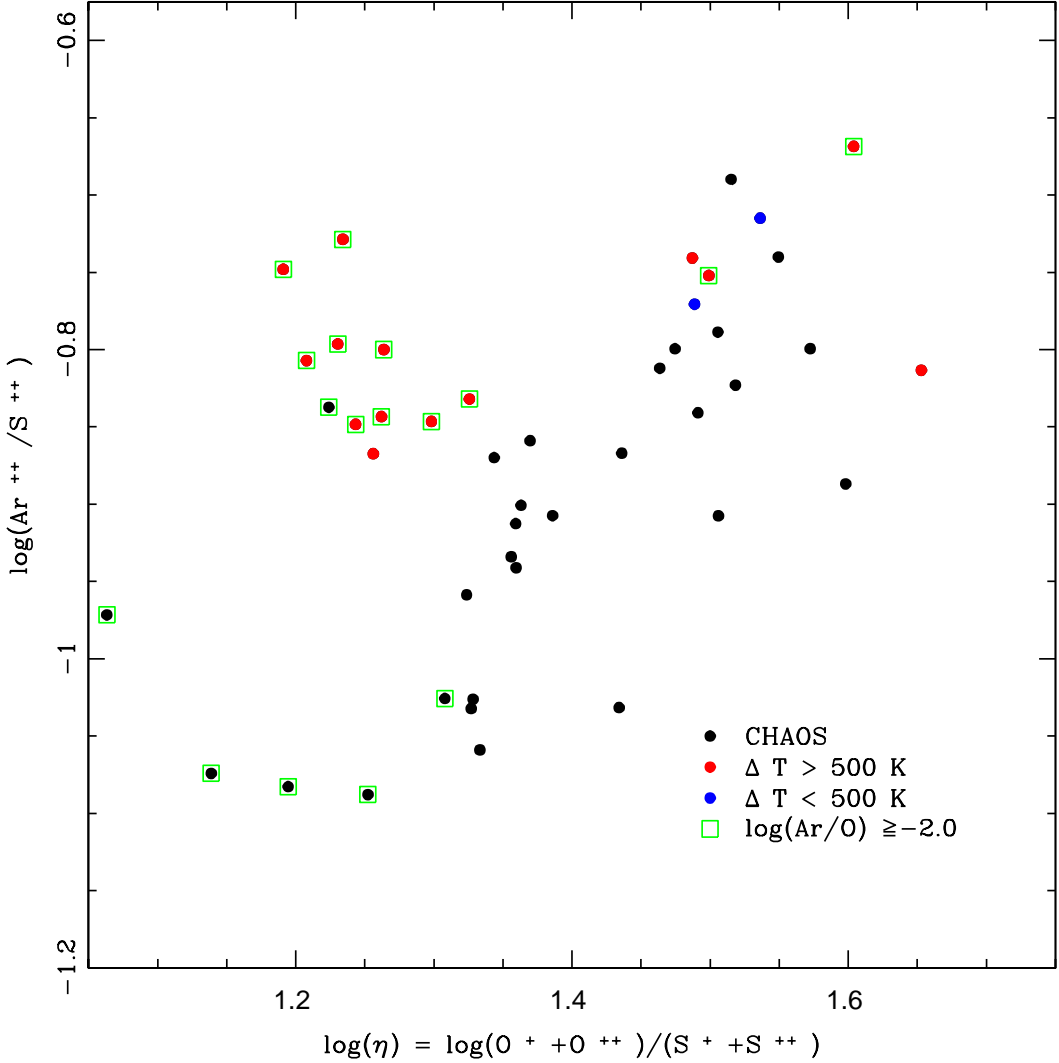}
\caption{Temperature sensitive indicators log(Ar$^{++}$/S$^{++}$) and log($\eta$) are plotted for NGC 628.
Colored points are used to indicate regions where both T[\ion{S}{3}] and T[\ion{O}{3}] were measured,
 where highly discrepant regions are colored red and non-discrepant regions are blue.
 Regions with only one temperature measurement are colored black.
 Log(Ar/O) outliers are outlined by green boxes.}
\label{fig13}
\end{figure}

 The ratio of [\ion{Ar}{3}] $\lambda7136$ to [\ion{S}{3}] $\lambda9069$ has been found to be insensitive to ionization
 structure, determined almost completely by the electron temperature and the ratio of the number of Ar and S ions 
 \citep{stevenson93}, and therefore offers a more controlled look at temperature discrepancies.
 In Figure~\ref{fig13} we plot Ar$^{++}$/S$^{++}$ ratio versus the $\eta$ parameter \citep{vilchezpagel88} for \ion{H}{2} regions in NGC~628 .
 The $\eta$ parameter ($\eta$ = log[(O$^+$/O$^{++}$)/(S$^+$/S$^{++}$)]) represents the hardness of the input ionizing spectrum,
 and is a sensitive function of stellar effective temperature.
 This parameter pairing allows a more isolated temperature sensitive comparison and the implied result is striking.
 Most of the data follows a a linearly increasing relationship for log(Ar$^{++}$/S$^{++}$) vs log($\eta$),
 which is further confirmed by the non-discrepant T[\ion{S}{3}]-T[\ion{O}{3}] blue points that lie neatly along this trend.
 Further, most of the highly temperature discrepant red points are clear outliers.
 Some black points are also outliers, but they do not contain both temperature measurements required to assess temperature discrepancy. 
Based on this plot, Ar$^{++}$/S$^{++}$ is a useful diagnostic to identify extremely discrepant values of T[\ion{O}{3}].

\begin{figure}
\plotone{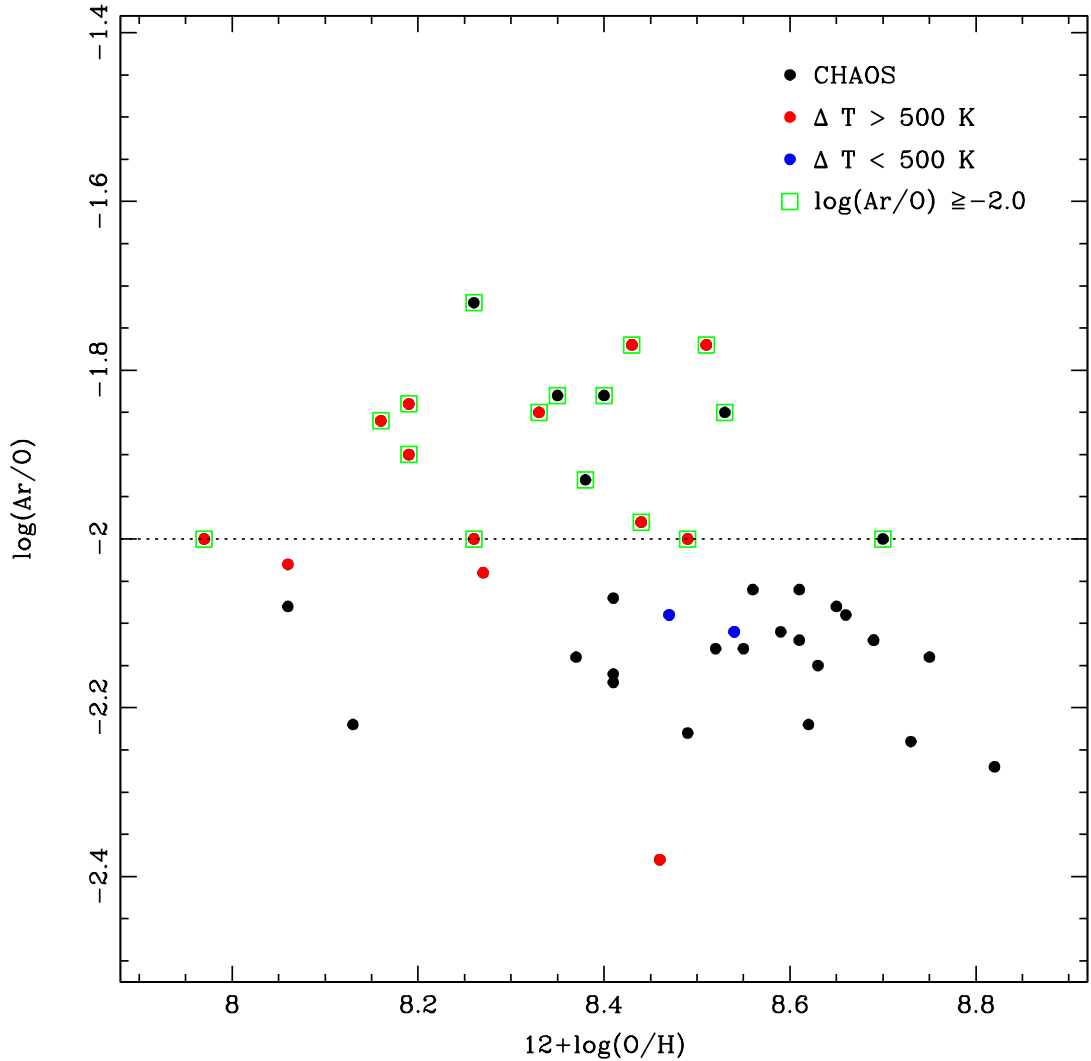}
\caption{The Ar/O ratio as a function of oxygen abundance is plotted for NGC628,
where T[\ion{O}{3}] is prioritized for the high ionization zone.
The expected $\alpha$-element trend is recovered for the majority of the sample, 
but the resulting Ar/O values are higher on average (log(Ar/O) $= -2.033$) than
the average value from our standard analysis (log(Ar/O) $= -2.219$; see Section~\ref{sec:alpha})
and the scatter greatly increased.
A number of high Ar/O outliers (log(Ar/O) $\ge -2.0$; above dotted line) are present as indicated by green boxes.
It is of interest that many of these outliers correspond to large discrepancies in T[\ion{S}{3}]-T[\ion{O}{3}].}
\label{fig14}
\end{figure}

A second temperature discrepancy indicator of interest can be found in the $\alpha$-element abundance ratios.
In general, $\alpha$-elements are thought to be produced primarily in massive stars and thus returned 
on similar time scales such that their ratio remains constant across oxygen abundance.
However, Ar$^{++}$ and O$^{++}$ lie in different ionization zones, and so if large temperature discrepancies exist, 
we can expect to see regions with outlying values of log(Ar/O).
Figure~\ref{fig14} demonstrates the usefulness of the Ar/O ratio as an indicator of temperature discrepancies.
Log(Ar/O) is plotted versus oxygen abundance, using the standard choice of temperature indicators 
in the abundance analysis, and for the majority of the sample the anticipated constant trend is recovered,
although the error-weighted average value is shifted up to log(Ar) $= -2.03$
and scatter is more than doubled ($\sigma_t = 0.186$; $\sigma_i = 0.172$).
Above the dominant horizontal distribution of non-discrepant temperature 
points are a number of points with log(Ar/O) $\ge -2$, whose separation is denoted by a dotted line.
Many of the Ar/O outliers, denoted by green boxes, correspond to large measured temperature discrepancies,
and appear to be systematically offset to higher Ar/O ratios from the main sample.
These outliers are consistent with the picture that overestimated [\ion{O}{3}] temperatures 
produce underestimated oxygen abundances and thus overestimated Ar/O ratios.
From these analyses, Ar has emerged as a possibly important gauge of temperature validity, 
and will be revisited in a more thorough analysis in a future paper.

\clearpage

\bibliography{mybib}{}

\clearpage


\end{document}